\documentclass[%
reprint,
superscriptaddress,
 amsmath,amssymb,
 showkeys,
prb,
floatfix,
]{revtex4-2}

\usepackage{graphicx}
\usepackage{dcolumn}
\usepackage{bm}
\usepackage{hyperref}
\usepackage[mathlines]{lineno}
\usepackage{gensymb}
\usepackage{subfigure}
\usepackage{xcolor}

\usepackage{hhline}

\begin{document}

\title{Phase Transition of Two-Dimensional Ferroelectric and Paraelectric Ga$_{2}$O$_{3}$ Monolayer: A Density Functional Theory and Machine-Learning Study}

\author{Junlei Zhao} \email{zhaojl@sustech.edu.cn}
\affiliation{Department of Electrical and Electronic Engineering, Southern University of Science and Technology, Shenzhen 518055, China}
\author{Jesper Byggm\"astar} \email{jesper.byggmastar@helsinki.fi}
\affiliation{Department of Physics, University of Helsinki, P.O. Box 43, FI-00014, Finland}
\author{Zhaofu Zhang} \email{zz389@cam.ac.uk}
\affiliation{Department of Engineering, University of Cambridge, Cambridge CB2 1PZ, United Kingdom}
\author{Flyura Djurabekova} \email{flyura.djurabekova@helsinki.fi}
\affiliation{Department of Physics, University of Helsinki, P.O. Box 43, FI-00014, Finland}
\affiliation{Helsinki Institute of Physics, University of Helsinki, P.O. Box 43, FI-00014, Finland}
\author{Kai Nordlund} \email{kai.nordlund@helsinki.fi}
\affiliation{Department of Physics, University of Helsinki, P.O. Box 43, FI-00014, Finland}
\affiliation{Helsinki Institute of Physics, University of Helsinki, P.O. Box 43, FI-00014, Finland}
\author{Mengyuan Hua} \email{huamy@sustech.edu.cn}
\affiliation{Department of Electrical and Electronic Engineering, Southern University of Science and Technology, Shenzhen 518055, China}


\begin{abstract}

Ga$_{2}$O$_{3}$ is a wide-band-gap semiconductor of great interest for applications in electronics and optoelectronics.
Two-dimensional (2D) Ga$_{2}$O$_{3}$ synthesized from top-down or bottom-up processes can reveal brand new heterogeneous structures and promising applications. 
In this paper, we study phase transitions among three low-energy stable Ga$_{2}$O$_{3}$ monolayer configurations using density functional theory and a newly developed machine-learning Gaussian approximation potential, together with solid-state nudged elastic band calculations.
Kinetic minimum energy paths involving direct atomic jump as well as concerted layer motion are investigated.  
The low phase transition barriers indicate feasible tunability of the phase transition and orientation via strain engineering and external electric fields. Large-scale calculations using the newly trained machine-learning potential on the thermally activated single-atom jumps reveal the clear nucleation and growth processes of different domains. 
The results provide useful insights to future experimental synthesis and characterization of 2D Ga$_{2}$O$_{3}$ monolayers.
   
\end{abstract}

\maketitle

\section{Introduction}

Gallium oxide (Ga$_{2}$O$_{3}$) is a highly promising candidate for next-generation power electronic devices \cite{pearton2018a, tsao2018ultrawide}. Owing to its wide band gap ($\sim$ 4.8 eV) \cite{Orita2000Deep}, a small electron effective mass \cite{Peelaers2015Bri, Furthm2016Quasi}, and the transparency well into the ultra-violet (UV) band, Ga$_{2}$O$_{3}$ nanolayer demonstrated its utility, for example, as a novel optoelectronic material for low‐cost passivation and protection of atomically thin semiconductors \cite{wurdack2021ultrathin} and deep-UV sensors \cite{wang2021progresses}. As the most stable phase among five polymorphs (labeled as $\alpha$, $\beta$, $\gamma$, $\delta$ and $\epsilon$ analogous to alumina), $\beta$-Ga$_{2}$O$_{3}$ adopts a monoclinic crystal structure \cite{peelaers2017lack}. 

\begin{figure}[ht!]
 \centering
 \includegraphics[width=8cm]{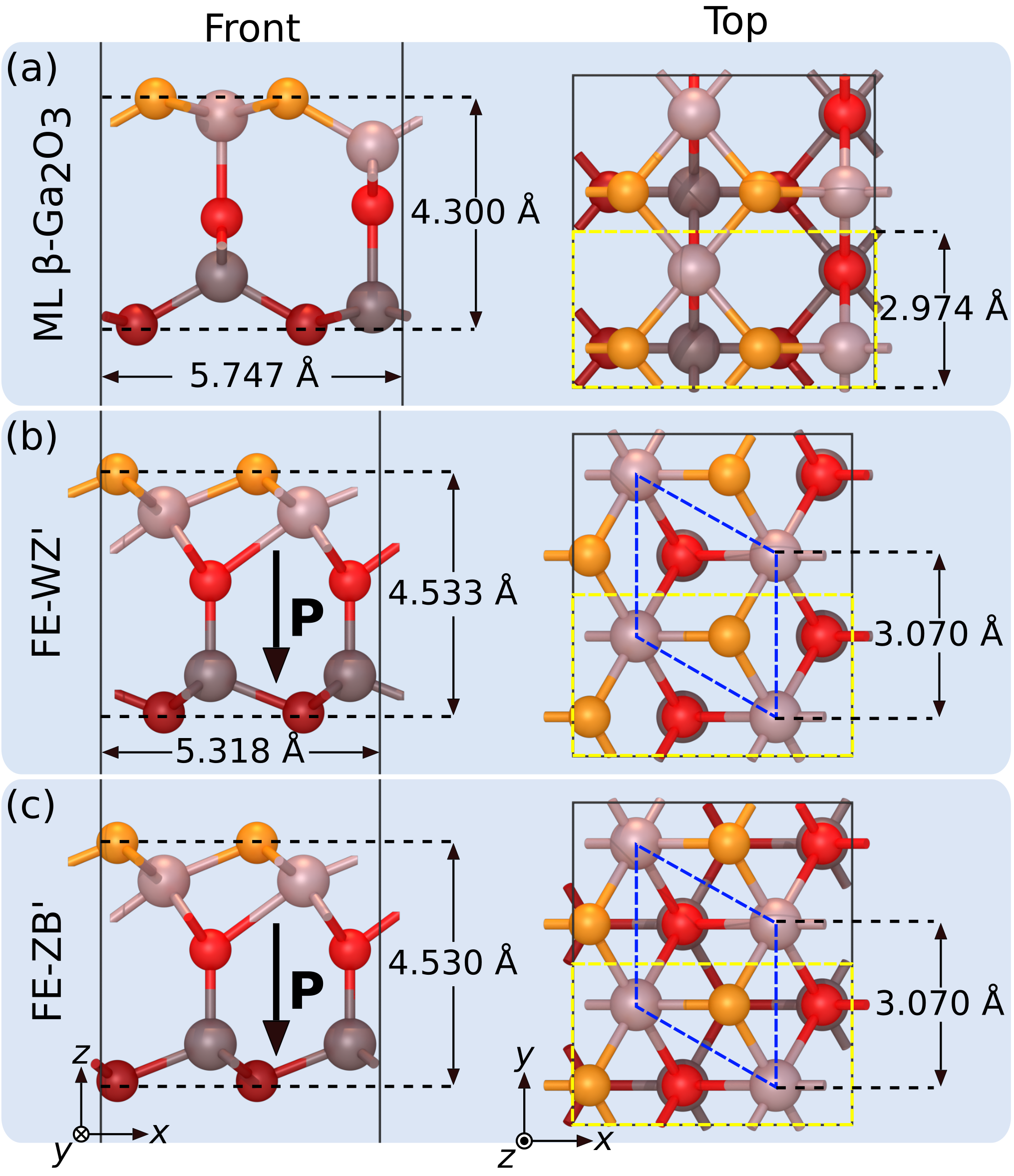}
 \caption{Atomic configurations of (a) ML-$\beta$, (b) FE-WZ' and (c) FE-ZB' phases. The colors of the Ga (big spheres) and O (small spheres) atoms are set differently to distinguish each atomic layer (three O layers and two Ga layers). The direction of spontaneous polarization is indicated by the black arrow in (b) and (c). The hexagonal primitive unit cells are marked by the blue parallelograms, while orthogonal cells used for CI-SSNEB are marked by the yellow rectangles, which is also the primitive cell for ML-$\beta$ phase.}
 \label{Layout}
\end{figure}

Although Ga$_{2}$O$_{3}$ is not a Van der Waals material, the 2D $\beta$-Ga$_{2}$O$_{3}$ can be mechanically exfoliated from the bulk material along the (100) direction, forming thin $\beta$-phase layers \cite{Zhou2017insulator, kwon2017tuning, Barman2019Mechanism}.  
Quasi-2D Ga$_{2}$O$_{3}$ thin films have been mechanically exfoliated with the thickness of $\sim$ 20 to 100 nm \cite{Hwang2014High, kwon2017tuning, zhou2017high, Chen2019Ohmic}.
Beside the mechanical exfoliation, the 2D Ga$_{2}$O$_{3}$ nanolayer can be synthesized bottom-up via epitaxial growth methods, such as atomic layer deposition (ALD) \cite{chandiran2012subnanometer}, metal organic chemical vapor deposition (MOCVD) \cite{ZHANG2020157810} and liquid metal-based reaction \cite{zavabeti2017liquid}.
It is also reported that the 2D $\beta$-Ga$_{2}$O$_{3}$ lacks quantum confinement, in a sense that the band gap and electron effective mass do not change significantly for the 2D thin layer compared to its bulk counterpart \cite{peelaers2017lack}. 
Ideally, a monolayer (ML) $\beta$-Ga$_{2}$O$_{3}$ (abbreviated as ML-$\beta$ in the later text) can be constructed by cutting half of the conventional unit cell of $\beta$-Ga$_{2}$O$_{3}$ from the most stable (100) plane. 
The ML-$\beta$ phase (Fig. \ref{Layout}a) has the centrosymmetric structure with the space group $P2/m$ (space group No. 10), unlike its bulk counterpart whose space group is $C2/m$ (space group No. 12). 
Moreover, ML-$\beta$ phase is paraelectric with no intrinsic out-of-plane dipole moment.  
For very thin nanolayers grown from deposition, the facet orientation of the substrate will strongly confine the ordering of the initial epitaxy layers at the interface. 
Therefore, besides the amorphous phase, the metastable crystallized phases can be expected to be seen during the early stage of the growth on highly ordered facets of substrates, such as the (111) surface of face-centered-cubic metal \cite{wang2019epitaxial, dong2020the, chen2020wafer, zhang2021hexagonal}.

Recently, two Van der Waals 2D ferroelectric (FE) configurations of Ga$_{2}$O$_{3}$ were discovered in our previous studies \cite{liao2020tunable} by adopting a class of stable single-layer configurations based on III$_{2}$-VI$_{3}$ compounds \cite{ding2017prediction, xiao2018intrinsic, zhao2018two, fu2018intrinsic} with an asymmetric quintuple-atomic-layer configuration. 
One Ga$_{2}$O$_{3}$ layer consists of five triangular atomic lattices stacked in the sequence O-Ga-O-Ga-O and belongs to $R3m$ space group (space group No. 160). 
Depending on stacking type of the fifth O layer, two almost energetically degenerate configurations are named as ferroelectric Wurtzite (FE-WZ') and ferroelectric Zinc Blende (FE-ZB') \cite{ding2017prediction, liao2020tunable}, as shown in Fig. \ref{Layout}b and c. 
Compared to the ML-$\beta$ phase, the FE-WZ' and FE-ZB' phases have higher energies of 0.430 eV per 10-atom orthogonal cell (see Fig. \ref{Layout}), while the FE-WZ' phase is marginally, 4$\times$10$^{-3}$ eV per 5-atom hexagonal cell, higher in energy than FE-ZB'. 
It is well known that metastable structures can appear in as-grown products due to kinetic trapping effect of the synthesizing conditions \cite{zhao2016formation, vernieres2019site}.
Therefore, the relative stability and phase transition path is of great interest and can provide useful insights on the bottom-up synthesis methods as well as possible post-processing and control of the structure via strain engineering or external electric field. 

In this work, we invoke a multiscale computational approach to explore polarization reversal transition and phase transition pathways among the ferro- and para-electric monolayers using well established density functional theory (DFT) and climbing-image (solid-state) nudged elastic band (CI-(SS)NEB) methods. By analyzing the energetics of potential transient configurations between stable and metastable states, deeper understanding can be gained towards eventual successful synthesis of the structure in experiment. Furthermore, we develop a Gaussian approximation potential (GAP) to employ CI-(SS)NEB calculations of the 2D Ga$_{2}$O$_{3}$ phases on a larger scale. 

The paper is divided into the following sections. In section II, the detailed methodology of the computational techniques are introduced. In section III-A, we focus on the polarization reversal transition of FE-ZB' and FE-WZ' configurations. In section III-B, we investigate the domain wall motions between two oppositely polarized FE-ZB' domains as well as FE-WZ' against FE-ZB' domains. In section III-C, we study the solid-state phase transition paths from FE-ZB' and FE-WZ' to the ML-$\beta$ phase. In section III-D, we employ the GAP to calculate the transition barriers during the nucleation process of the oppositely polarized FE-ZB' phases and the solid-state FE-ZB'$\rightarrow$ML-$\beta$ phase transition. We conclude with discussion and comparison of the computational results with the recent experimental findings \cite{zavabeti2017liquid, wurdack2021ultrathin} in section IV.

\section{Computational Methods}

\subsection{DFT details}

The DFT calculations were conducted using the Vienna Ab-initio Simulation Package (VASP) \cite{kresse1993ab, kresse1996efficiency}, employing the projected augmented-wave (PAW) method \cite{PAW}.
To account for Van der Waals interactions, we used Grimme's long-range dispersion correction (DFT-D3) \cite{DFTVDW}.
The dipole correction was considered on the potential and forces along the $z$ axis throughout the whole calculations.
In the DFT calculations, the electronic states were expended in plane-wave basis sets with an energy cutoff of 700 eV.
The Brillouin zone was sampled with a $\Gamma$-centered k-mesh grid with spacing 0.2 \r A$^{-1}$ which was equivalent to a dense 11$\times$11$\times$2 grid for a hexagonal 3.07 \r A$\times$ 3.07 \r A$\times$25 \r A unit cell. 
Gaussian smearing with a width of 0.03 eV was used to describe the partial occupancies of the electronic states. 
The detailed convergence tests on the plane-wave energy cutoff and the k-mesh grid are attached in the Electronic Supplementary Information (ESI) Fig. S1.
We chose 10$^{-6}$ eV and 5$\times$10$^{-3}$ eV/\r A as the energy and force convergence criteria for the optimization of the electronic and ionic structures, respectively. 
The Perdew-Burke-Ernzerhof version of the generalized gradient approximation (GGA-PBE) \cite{perdew1996generalized} was used for the initial configuration optimization and CI-(SS)NEB calculations. We note the well known fact that the approximated functional such as PBE used here, can underestimate reaction barrier due to the intrinsic delocalization error \cite{cohen2008insights, momeni2020multiscale}. However, this possible error is expected to scale systematically, and hence should not alter the overall picture of the transition pathways.  

The in-plane lattice constants of FE-ZB' phase was reported by the previous study in Ref. \cite{liao2020tunable}. In this work, we finetuned the in-plane lattice constants of the FE-ZB' and FE-WZ' phases with more accurate configuration relaxation (see the ESI Fig. S2 for detail). We note that this is an essential step for the NEB calculations, as otherwise it could lead to the problematic result that an intermediate image may pass through the real minima instead of the fixed terminal images. The lattice constants ($a$ and $b$) of the ML $\beta$-phase were obtained by fully relaxing a bulk-cut configuration in an orthogonal cell. The lattice constants of the FE-WZ' and FE-ZB' phases were relaxed in a in-plane hexagonal cell. A vacuum layer with a thickness of 20 \r A was used to avoid interactions between the periodic images in the $z$ direction throughout the calculations. We note that the in-plane lattice constants of the ML-$\beta$ phase are smaller than the ones of the bulk $\beta$-Ga$_{2}$O$_{3}$ ($a = 3.08$ \r A, $b = 5.92$ \r A). The in-plane lattice parameters of the FE-WZ' and FE-ZB' phases are the same, while a very minor change is seen in out-of-plane thickness as shown in Fig. \ref{Layout}.

\subsection{Nudged elastic band calculations} 
 
Standard CI-NEB \cite{jonsson1998nudged, henkelman2000a, henkelman2000improved} with the fixed lattice vectors was used to calculate the activation barriers of polarization reversal, FE-ZB'$\rightarrow$FE-WZ' transitions, and domain wall motions, while CI-SSNEB \cite{SSNEB} was used to calculate the phase transitions of FE-ZB'$\rightarrow$ML-$\beta$ and FE-WZ'$\rightarrow$ML-$\beta$ paths in order to account the primarily lattice changes. In the standard CI-NEB, only the changes in the internal Cartesian positions of the atoms are included to calculate spring forces between images, while CI-SSNEB couples the cell and atomic variables by concatenating the scaled cell strain and the changes in atomic positions.  

Because of the different symmetries between the primitive hexagonal and orthogonal cells, we adopted the initial images of FE-ZB' and FE-WZ' in 10-atom orthogonal cells as marked by the yellow rectangles in Fig. \ref{Layout}b and c. In this way, the solid-state transition can be decomposed into two in-plane orthogonal cell strains and stresses and internal atomic forces. With the lattice vector fixed at 25 \r A in the $z$ direction, the out-of-plane stress was relaxed depending on the internal atomic forces. The calculations were stopped after the force acting on the saddle point image fell below 0.01 eV/\r A. We note that the energy changes at the saddle point were converged below 10$^{-3}$ eV after 150 iterations at maximum. The standard CI-NEB calculations were done with the Atomic Simulation Environment (ASE) framework \cite{ASE}. The CI-SSNEB calculations were using the TSASE library \cite{TSASE}. OVITO was used for the visualization of the atomic configurations \cite{stukowski2010visualization}.  

\subsection{Construction of a Gaussian approximation potential for 2D phases}   

In order to overcome the temporal and spatial limits of \textit{ab-initio} methods, we constructed a kernel-based machine-learning Gaussian approximation potential \cite{GAP, SOAP}. Unlike other existing machine-learning potentials which are purely developed for studying thermal properties of bulk $\beta$-Ga$_{2}$O$_{3}$ \cite{liu2020machine, li2020a}, we explicitly trained our GAP for both bulk and 2D structures. The training and testing database are generated from \textit{ab-initio} molecular dynamics (AIMD) simulations. The AIMD simulations for the 2D phases were performed with the same setting mentioned in Section IIA for sampling the electronic system. The simulation cells of ML-$\beta$, FE-ZB' and FE-WZ' consists of the 2$\times$4$\times$1 orthogonal supercell ($\sim$11 \r A$\times$12 \r A$\times$25 \r A) with 80 atoms. To improve transferability of the GAP for modeling primarily the bulk structures, we included one third of the configurations from the bulk $\beta$ phase to the training and testing database. For the bulk $\beta$ phase, the 1$\times$4$\times$2 monoclinic supercell ($\sim$12 \r A$\times$12 \r A$\times$12 \r A) with 160 atoms are used. For each configuration (three 2D and bulk $\beta$ phases), the simulation cells are scaled with uniform strain on non-vacuum lattice constant from -4\% to 4\% with the step of 2\%. The AIMD simulations are run at 100/300/600/900K for all strains for 1 ps with a time step of 2 fs. Additionally, the FCC phase (see Fig. \ref{WZ_ZB}f), the intermediate minima (V1, V2, V3) and the five saddle-points configurations from the CI-SSNEB (see Fig. \ref{ZB_beta_WZ}) calculations are taken for constructing 2$\times$4$\times$1 supercells with 80 atoms. 20 independent single-point DFT calculations on each of these supercells are computed, with small normally distributed random displacements added to the atomic positions to create unique and diverse local atomic environments with nonzero forces. The single Ga/O atom in vacuum and Ga-Ga, Ga-O, O-O dimer systems are included for reference energy and repulsive forces at close atomic distances. In total, 932 configurations with 81,522 atomic environments are in the training database and 743 configurations with 71,200 atomic environments in the testing database. The total energies, atomic forces and virial stresses are stored for training and testing. 

Both the two-body descriptor $\textbf{q}_\mathrm{2b}$ \cite{NEW2B} and the many-body smooth overlap of atomic positions (SOAP) descriptor $\textbf{q}_\mathrm{mb}$ \cite{SOAP} are used. The total energy of $N$ atoms is then given by 
\begin{equation} \label{Eq1}
\begin{split}
E_\mathrm{tot.} & = \delta_\mathrm{2b}^{2} \sum_{i}^{N_\mathrm{pairs}} \sum_{s}^{M_\mathrm{2b}} \alpha_{s,\mathrm{2b}}K_\mathrm{2b}(\textbf{q}_{i,\mathrm{2b}}, \textbf{q}_{s,\mathrm{2b}}) \\
& + \delta_\mathrm{mb}^{2} \sum_{i}^{N} \sum_{s}^{M_\mathrm{mb}} \alpha_{s,\mathrm{mb}}K_\mathrm{mb}(\textbf{q}_{i,\mathrm{mb}}, \textbf{q}_{s,\mathrm{mb}}),
\end{split}
\end{equation}
where $\delta_\mathrm{2b}$ and $\delta_\mathrm{mb}$ are the pre-factors of the Gaussian process, $\alpha_\mathrm{2b}$ and $\alpha_\mathrm{mb}$ are the regression coefficient vectors to be fitted during the training, $K_\mathrm{2b}$ is the squared exponential kernel and $K_\mathrm{mb}$ is the dot product kernel. The detailed hyper-parameters used for constructing the descriptors, the kernel functions, and training the GAP are summarized in the ESI Table S1. A detailed analysis of the accuracy of the GAP can be found in the ESI Fig. S6. For more details about the construction of Gaussian approximation potentials, we refer the reader to Refs. \cite{GAP, SOAP, GAP2015}. The training processes are performed using the QUantum mechanics and Interatomic Potentials (QUIP) package \cite{QUIP}. 

\section{Results}
    
\subsection{Polarization reversal transition}

\begin{figure*}[ht!]
\centering
\includegraphics[width=16cm]{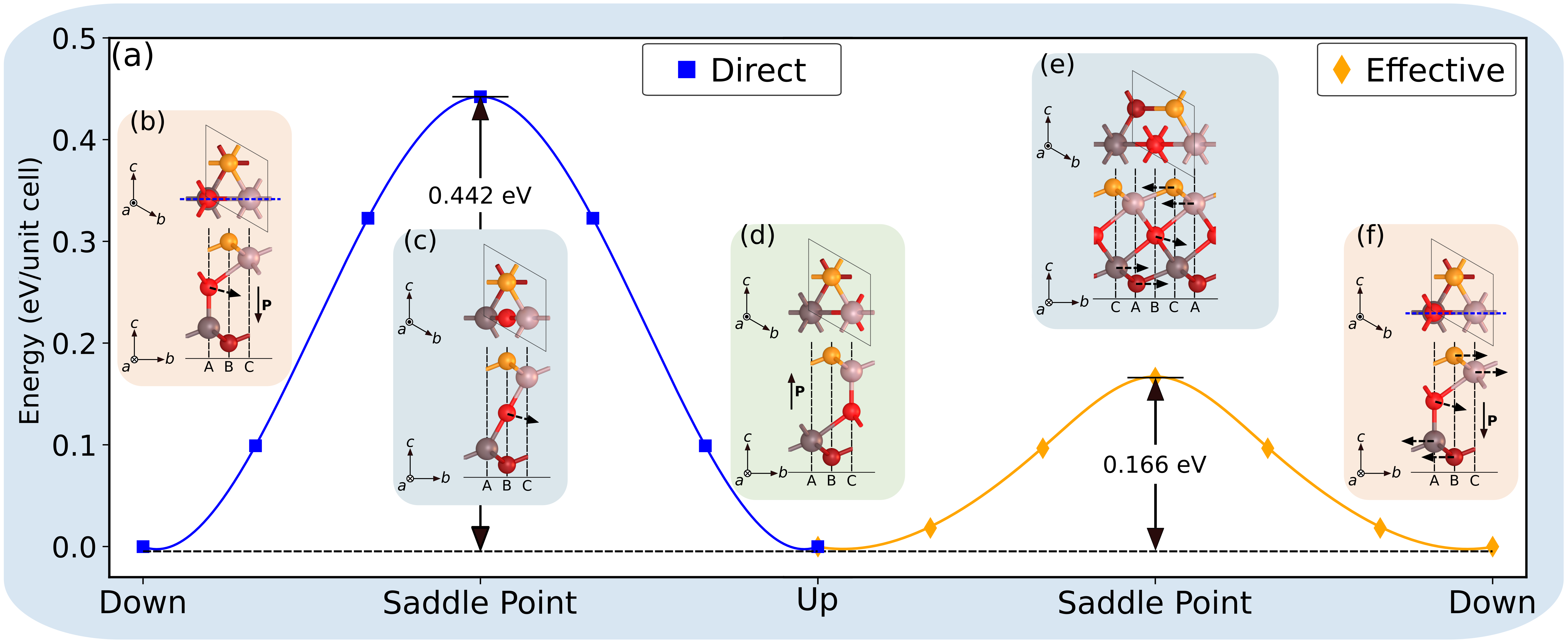}
\caption{Polarization reversal transition of FE-ZB' phase: Evolution of the energy of monolayer Ga$_{2}$O$_{3}$ in the FE-ZB' phase transforming from the state with the electric polarization pointing downward (b, f) to the state with the electric polarization pointing upward (d) via a direct shifting process (b$\rightarrow$c$\rightarrow$d): the O atoms in the central layer laterally shift from the A to C sites. The most effective kinetic pathway (f$\rightarrow$e$\rightarrow$d) involves a three-step concerted mechanism, as detailed in the main text. The dashed black arrows attached to atoms indicate the directions of atomic motion during the polarization reversal processes along the plane labeled with the blue dashed line in (b, f). The energy differences are shown for a 5-atom hexagonal unit cell.}
\label{ZB_filp}
\end{figure*}  

Polarization reversal transition is a special type of structural phase transition in ferroelectric III$_{2}$VI$_{3}$ compounds, which could lead to potential interesting applications such as bit flipping in data storage \cite{Scott2007ferro}. First, using standard CI-NEB we investigated the most effective kinetic pathway of the polarization reversal transition of the FE-ZB' and FE-WZ' phases without an external electric field. Second, in order to illustrate the tunability of the reversal transition, the effect of the out-of-plane external field on the reduction of the transition barrier was studied. 

For the FE-ZB' phase, as shown in Fig. \ref{ZB_filp}a, the transition barrier for the direct jump within the central layer of O atoms from site A to site C is 0.442 eV per unit cell (shown by one black arrow in Figs. \ref{ZB_filp}b$\rightarrow$c$\rightarrow$d), while an effective concerted transition process (shown by several black arrows in Figs. \ref{ZB_filp}f$\rightarrow$e$\rightarrow$d) has much lower barrier of 0.166 eV per unit cell. Initially the three upper atomic layers (one of Ga and two of O atoms) and two lower atomic layers (one of Ga and one of O atoms) move in-plane in opposite directions transforming the FE-ZB' phase into an unstable FCC structure as shown in Fig. \ref{ZB_filp}e. After that, the two uppermost layers of O and Ga atoms and the two lowest layers of Ga and O atoms reverse their motion and return back to their initial positions, while the central O layer finalizes its transition to the final site C. We note that the FCC phase for the Ga$_{2}$O$_{3}$ monolayered structure is unstable, which is unlike the other 2D III$_{2}$IV$_{3}$ materials, where this phase was found to be metastable \cite{ding2017prediction}. The single-atom jumps such are usually seen in thermally-activated processes, while the concerted atomic motions requires additional bias collectively acting on the system. Therefore, the consistently lower barriers of the concerted motions suggest the feasibility of tuning phase transition using external electric field. 

\begin{figure}[hb]
\centering
\includegraphics[width=8cm]{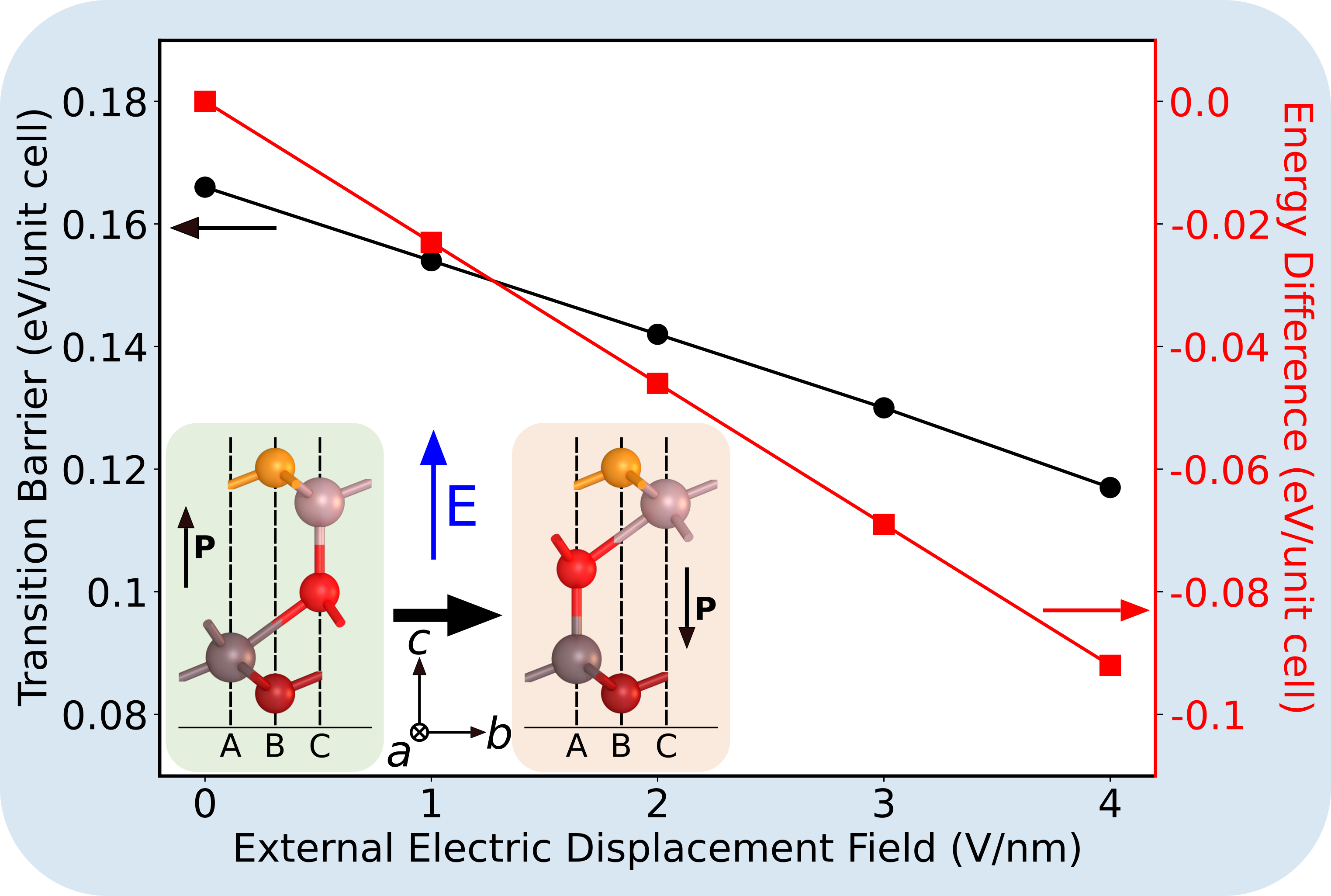}
\caption{The calculated transition barrier (black circles) and energy difference (red squares) between the initial and final states (the insets) in the electric polarization reversal process of FE-ZB' Ga$_{2}$O$_{3}$ via the concerted motion as illustrated in Fig. \ref{ZB_filp}f$\rightarrow$e$\rightarrow$d, plotted as a function of the external electric field applied in the out-of-plane direction. The applied external electric field is indicated by the blue arrow.}
\label{ZB_V_E}
\end{figure}  

As the second stage, we investigated the effect on the transition barrier of the external electric field applied along the normal to the substrate surface. As shown in Fig. \ref{ZB_V_E}, the energy difference between the two orientations as well as the transition barrier decrease linearly with the field strength increasing from 0 to 4 V/nm. We note that the energy barrier decreases much less rapidly than the energy difference (-0.12 and -0.23 eV per V/nm, respectively), indicating that the configuration at the saddle point exhibit less ferroelectric effect as expected, since the out-of-plane dipole moment in the FCC phase in the absence of an external electric field is zero due to symmetric stacking. The further comparison of the ionic configurations of the saddle points reveals a minor vertical off-center displacement of 0.015 \r A (0.054\%) of the central O layer under highest strength of the applied field of 4 V/nm. The detailed transition barriers are shown in the ESI Fig. S4. 

\begin{figure*}[ht!]
\centering
\includegraphics[width=16cm]{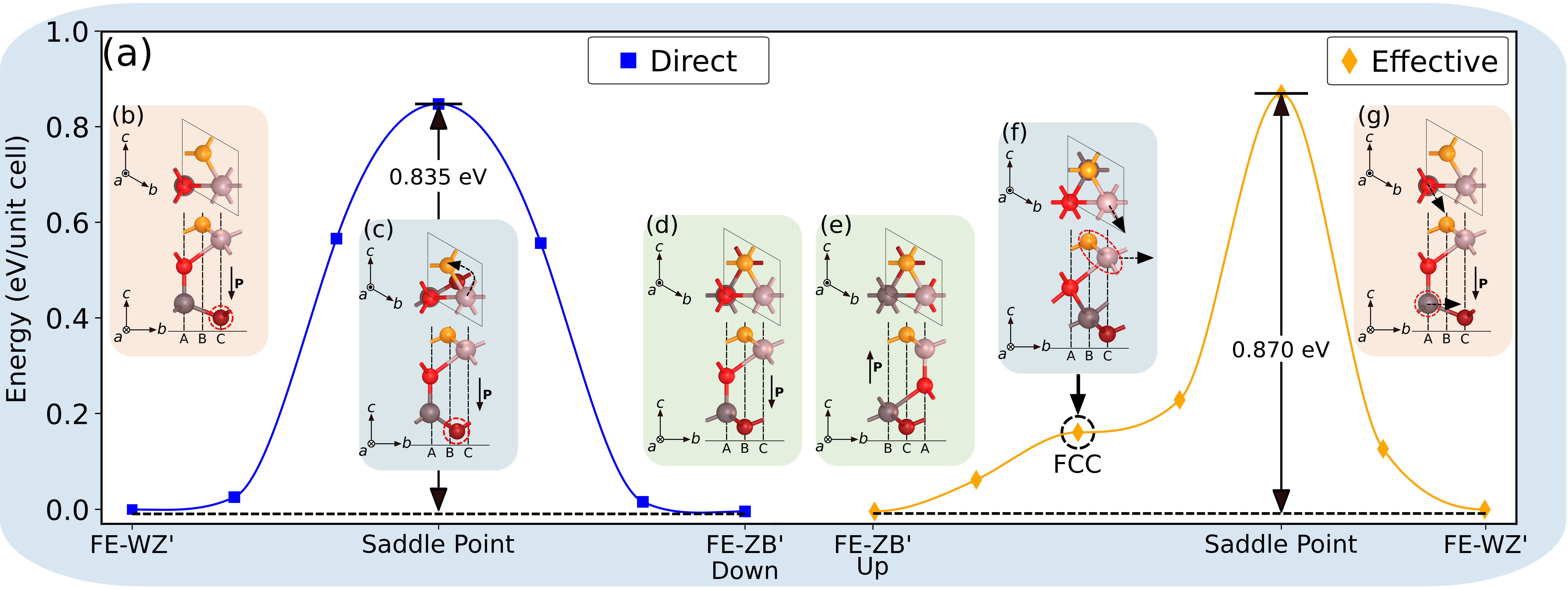}
\caption{FE-WZ'$\rightarrow$FE-ZB' transition. (a) Evolution of the energy of monolayer Ga$_{2}$O$_{3}$ transforming from FE-WZ' phase (b, g) to FE-ZB' phase (d, e). Left: a direct shifting process: the bottom O layer laterally shifts from the C to B sites. Right: a multi-stepped pathway involving the motion of the bottom Ga layer from A to B forming the FCC phase. The FCC phase can transform to FE-ZB' up or down with equal probability. The energy differences are shown for a 5-atom hexagonal unit cell.}
\label{WZ_ZB}
\end{figure*}  

We further focused on the phase transition path between FE-WZ' and FE-ZB' phases. As shown in Fig. \ref{WZ_ZB}, both the direct jump and the effective concerted movement of many atoms yield relatively high barriers (0.835 eV and 0.870 eV per unit cell, respectively). The direct jump of the bottom O layer from the stacking sites C to B does not change the polarization of the layer (Fig. \ref{WZ_ZB}b$\rightarrow$c$\rightarrow$d). However, the transition via the concerted movement of many atoms shown as Fig. \ref{WZ_ZB}g$\rightarrow$f$\rightarrow$e includes the non-polarized FCC configuration (Fig. \ref{WZ_ZB}f). This configuration is exactly the same as the one shown in Fig. \ref{ZB_filp}e, hence depending on the transition direction, the resulting orientation of the ZB-FE' phase can be either up or down. In Fig. \ref{WZ_ZB}e, the reverse polarization case is shown. 

\subsection{Domain wall motion}

\begin{figure*}[ht!]
\centering
\includegraphics[width=16cm]{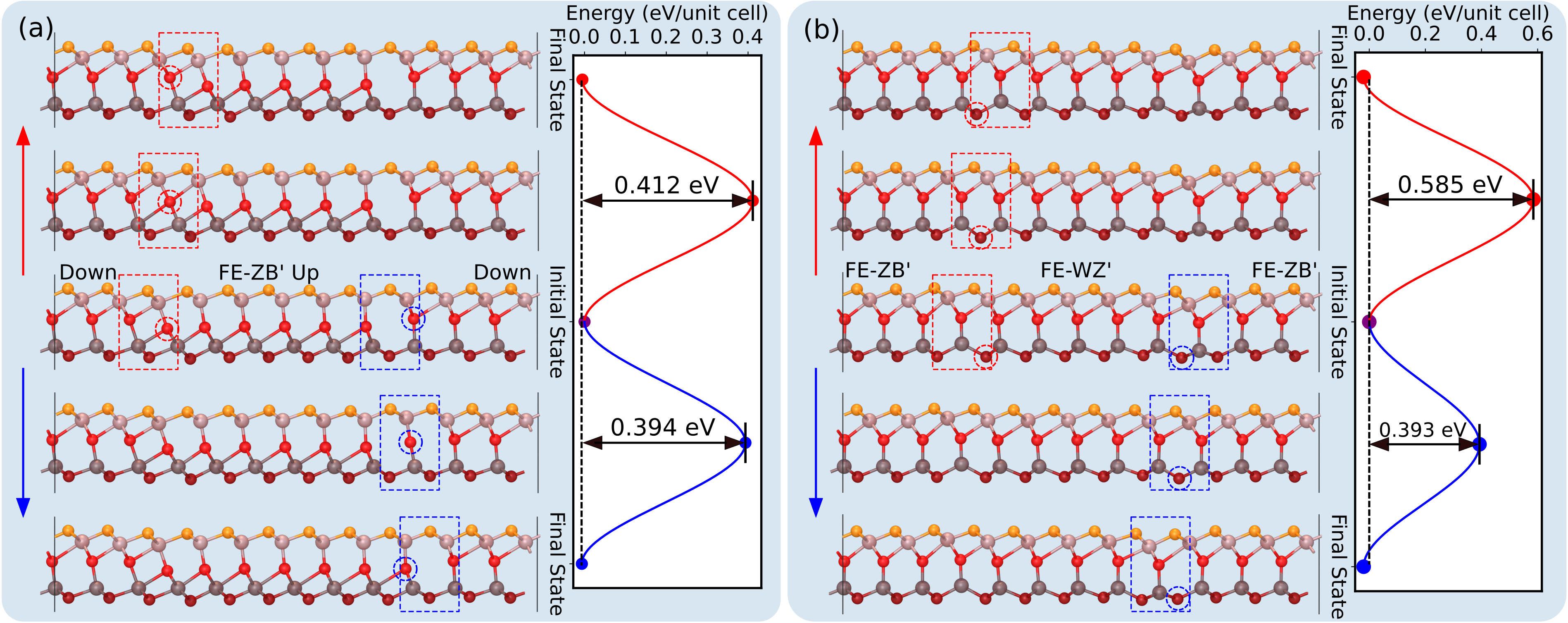}
\caption{(a) The domain wall motion of the polarization reversal process of the FE-ZB' phase with opposite electric polarization. (b) The domain wall motion of the ZB-WZ'$\rightarrow$FE-ZB' transition. The red and blue dashed boxes indicate the positions of the domain walls, and the dashed circles indicate the moving O atoms. The energy differences are shown for a 60-atom cell.}
\label{domain_all}
\end{figure*}  
 
While grown experimentally, the system may contain a number of different 0D and 1D defects, such as domain walls for 2D materials \cite{dawber2005physics}. Here, we further study the movement of two types of domain walls which facilitate the process of polarization reversal for the FE-ZB' phase. As shown in Fig. \ref{domain_all}a, we constructed a 12$\times$1 structure which contains two oppositely polarized domains by displacing six O atoms in the central layer vertically aligned to the top layer of Ga atoms. A large supercell was chosen to avoid the cross influence of the two domain walls. This structure after further relaxation to a local minimum was used as the initial configuration. Two types of the domain walls are labeled by the red and blue dashed boxes in the initial state of Fig. \ref{domain_all}a. 

It can be seen that the O atoms of the central layer at the domain walls are deformed from the perfect vertical alignment to the top or bottom Ga layers after the relaxation, due to the Coulomb repulsion between two close O atoms. The red and blue arrows on the left side of Fig. \ref{domain_all}a indicate the corresponding transitions of the two domain walls, where the circled O atoms of the central layer move to the vicinal polarization reversed sites. The transition barriers are 0.412 and 0.394 eV per unit cell which are slightly lower than the barrier of 0.442 eV per unit cell corresponding to the direct jump of the O atom from the central layer as discussed earlier (Fig. \ref{ZB_filp}a).  

The domain boundary between the FE-ZB' and FE-WZ' phases shown in Fig. \ref{domain_all}b were studied in a similar manner as Fig. \ref{domain_all}a, but with the displaced and moving O atoms of the bottom layer. The circled O atoms move from the stacking site of the top-layer Ga to the vicinal site of top-layer O. Here, the transition barriers of 0.585 and 0.393 eV per unit cell are significantly lower than the barrier of 0.835 eV of the direct jump of the bottom O layer (see Fig. \ref{WZ_ZB}). Moreover, the energy difference of 0.02 eV per unit cell between the initial and final state indicates that a larger fraction of  FE-ZB' phase is more likely to be seen in experiment.

Here we invoke the Arrhenius equation \cite{Arrheniuseq} to predict the transition rate, $r_{d}=\nu\exp{(-E_{a}/k_{B}T)}$, where $\nu$ is the attempt frequency in the order of 10$^{13}$ Hz, $k_{B}$ is the Boltzmann constant, $T$ is the temperature and $E_{a}$ is the transition barrier. At 300 K, the domain wall motion with the lower barrier of 0.393 eV moves more than 1,600 times faster than the ones with the barrier of 0.585 eV. For a 1-$\mu$m domain wall, the fastest propagation speed is about 0.68 nm/s at 300 K, which can be expected to be directly measured by microscopic characterization methods.    

\subsection{Solid-state phase transition path: FE-ZB' and FE-WZ' to ML-$\beta$ phase}

\begin{figure*}[ht!]
\centering
\includegraphics[width=16cm]{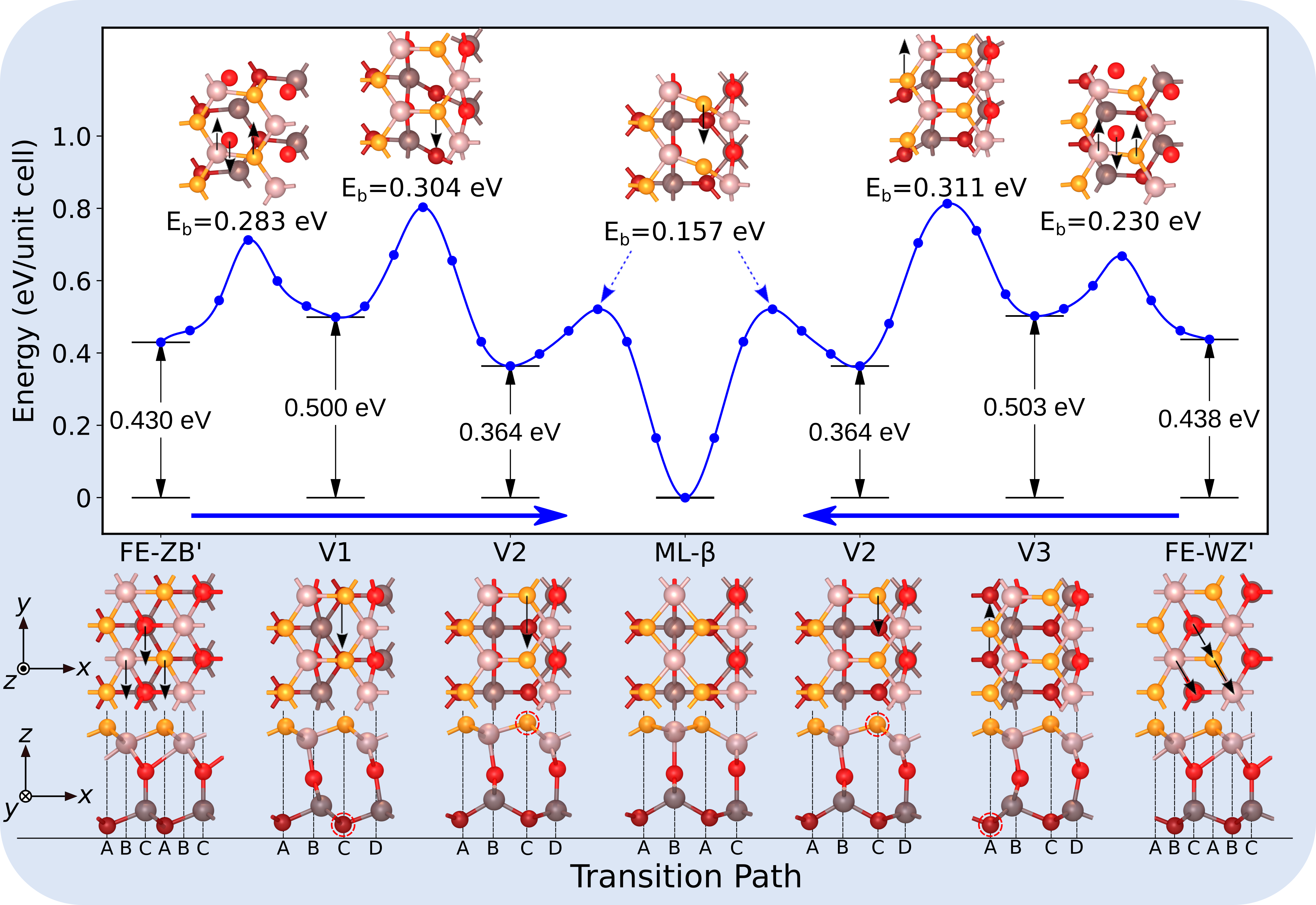}
\caption{Solid-state phase transition pathways. Upper: Evolution of the energy of phase transition from FE-ZB' and FE-WZ' phase to the ML-$\beta$ phase. The energy differences are shown for a 10-atom orthogonal unit cell. The intermediate transition barriers $E_{b}$ are with respect to the initial states, e.g., indicated by the black arrows at the bottom. Lower: the corresponding atomic configurations of the intermediate stages. The black arrows associated to the atoms indicate the moving direction. The red dashed circles label the moving O atoms. See also the ESI Movies S1 and S2 for the full FE-ZB'$\rightarrow$ML-$\beta$ and FE-WZ'$\rightarrow$ML-$\beta$ transitions, respectively.}
\label{ZB_beta_WZ}
\end{figure*}  

Using the CI-SSNEB method to account for the lattice changes, we explored the phase transition paths from FE-ZB' and FE-WZ' to the ML-$\beta$ phase. The complex paths with multiple local minima are seen in Fig. \ref{ZB_beta_WZ}. For each transition between two neighboring minima, we ran an independent CI-SSNEB calculation to determine accurately the position of a saddle point between them. Three intermediate minima are identified and named V1, V2 and V3 (V short for "Valley") as shown in Fig \ref{ZB_beta_WZ}. For the FE-ZB'$\rightarrow$ML-$\beta$ transition, the first step of the transition (FE-ZB'$\rightarrow$V1) involves a concerted movement of the top O-Ga-O layer in the $y$ direction, as indicated by the black arrows in Fig. \ref{ZB_beta_WZ}. At the saddle point, the top two layers of O and Ga atoms move back to their original stacking position, while the central O layer continues transition beneath the upper Ga layer towards the V1 state. Next, the bottom O layer migrates in the $y$ direction to ML-$\beta$ stacking. Finally, the top O layer moves in the $y$ direction to complete the ML-$\beta$ phase, as shown as V1$\rightarrow$V2$\rightarrow$ML-$\beta$ path in Fig. \ref{ZB_beta_WZ}. Similarly, three intermediate steps are identified in FE-WZ'$\rightarrow$ML-$\beta$ transition as well. The third step is the same movement from V2 to ML-$\beta$ phase with the barrier of 0.157 eV. The clear solid-state lattice transformations for both paths happen during FE-ZB'$\rightarrow$V1 and FE-WZ'$\rightarrow$V3 steps, where the stacking symmetry of the top and bottom Ga layers are changed. The following movements of the top and bottom O layers are driven mainly by atomic migration with minor changes in the in-plane lattice vectors of the cells.

\subsection{GAP-predicted phase transition}

\begin{figure*}[ht!]
\centering
\includegraphics[width=16cm]{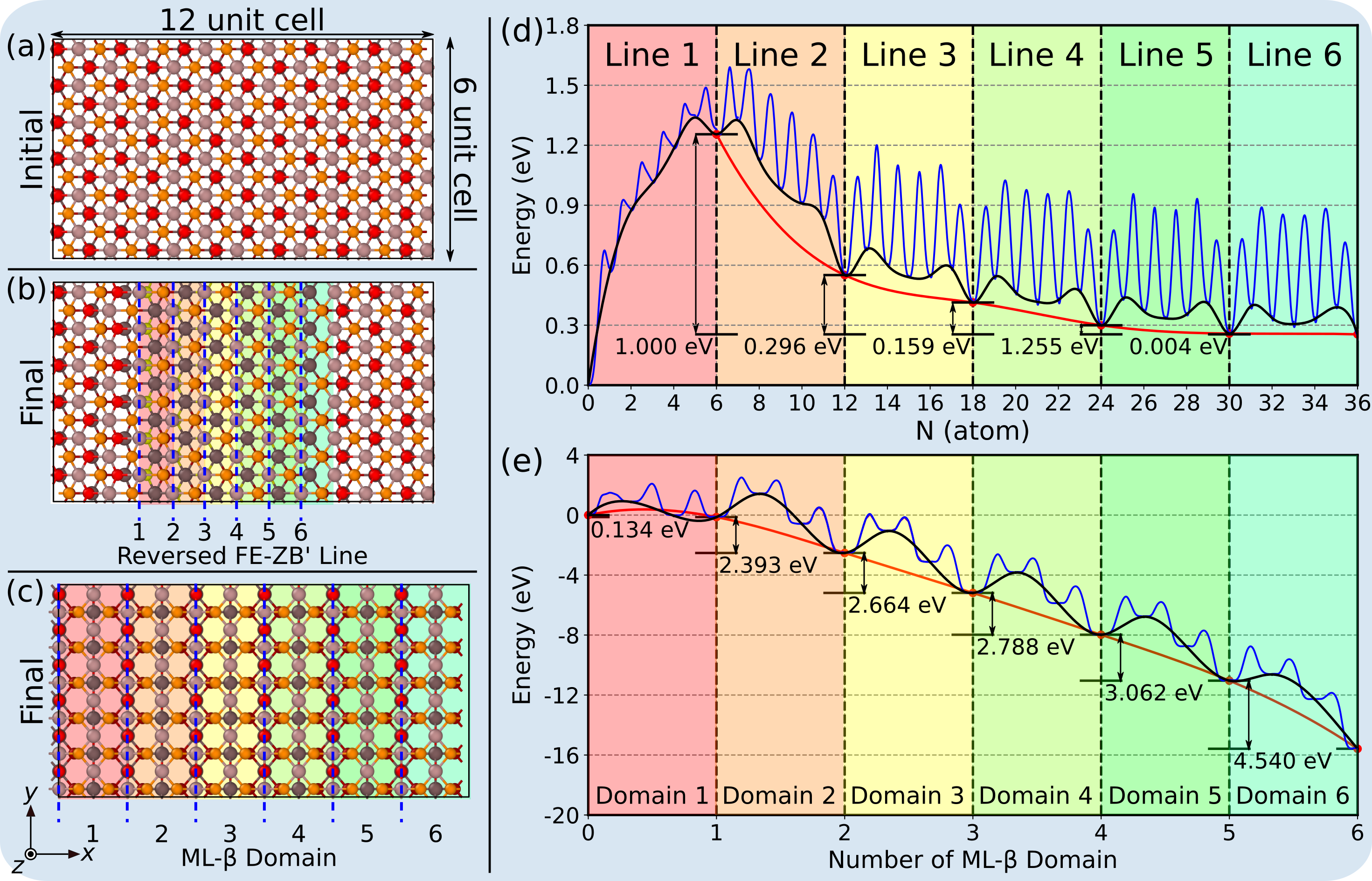}
\caption{(a) Initial 12$\times$6 FE-ZB' supercell; (b, d) the polarization reversed domain of the FE-ZB' phase; (c, e) the FE-ZB'$\rightarrow$ML-$\beta$ phase transition.
The blue curves show the transition barriers of single-step jumps which in (d) are the single-atom jumps and in (e) are the single-line jumps; the black curves illustrates the local minima and the red curves indicates the reduction in potential energy owing to the domain wall separation. The vertical dashed lines label the transition where a whole line/domain is completed. The energy differences are shown for a 360-atom cell. See also the ESI Movies S3 and S4 for the full processes, respectively.}
\label{MLdomain1}
\end{figure*}

So far, we have constrained the transition calculations within a limited size cell. This constraint may lead to an artifact showing a single atom migration as a collective movement of the atoms across the periodic boundaries. However, the CI-(SS)NEB calculations require multi-image force convergence up to hundreds of iterations of self-consistent calculations, thus cannot be afforded for a large supercell using {\it ab-initio} methods. To overcome this limit, we performed a large-scale NEB calculations using the GAP potential \cite{GAP2021} specifically trained to describe the 2D Ga$_{2}$O$_{3}$ phases with high accuracy of energies, atomic forces and cell stresses (see the ESI Fig. S6). We further constructed a full 2D 12$\times$6 FE-ZB' supercell with 360 atoms, as shown in Fig. \ref{MLdomain1}a. By moving the central-layer O atoms one at a time, the nucleation and growth of the domain with the reversed polarization can be calculated as shown in Fig. \ref{MLdomain1}d and the final state in Fig. \ref{MLdomain1}b. The full transition process can be found in the ESI Movie S3. Every six moving central-layer O atoms with the same initial $y$ positions are grouped and indexed as 0-5, 6-11 and 12-17, ..., which can complete a full line of the reversed domain. The initial nucleation of a full line of reversed domain leads to an increase of 1.27 eV in energy (``Line 1" region in Fig. \ref{MLdomain1}d), while the following growth reveals a gradual energy decrease which is caused by the separation of the two domain walls (see Fig. \ref{domain_all}a). Within a line growth, a periodic feature of the energy evolution can be seen as the displacement of the front atoms (i.e., atoms 0, 6, 12, 18...) always lead to increase in the potential energy, while the closing atoms (the last atoms completing the concerted movement, i.e. atoms 5, 11, 17) bring the system to more stable local minima. 

The blue spiked curves present the transition barriers of the single-atom jumps, which range from 0.15 to 0.58 eV. The variation originates from the different local atomic environments, such as domain width and starting or completing a domain wall. Comparison of the barriers for the overall shift -- 0.44 eV in Fig. \ref{ZB_filp}a, -- and for moving a domain wall -- 0.39 eV in Fig. \ref{domain_all}a where the whole line of the central-layer O atoms moves concertedly, -- the individual barriers reveal more clearly the effect of the initial nucleation of the polarization reversed FE-ZB' domain. We see that the barriers for the forward transitions are always much higher than those of the backward transitions in ``Line 1" region. Once the ``Line 1" domain is completed, the growth becomes energetically more favorable in ``Line 2" region where the forward barriers are smaller than the barriers of the reverse transitions. From ``Line 3" to ``Line 4", the single-atom barriers are simultaneously affected by the propagation within the individual lines and during the domain wall separation. Eventually when the two domain walls are far apart, the barriers depend on whether the moving atom is a starting or a completing the line atom as seen in Region ``Line 5" and ``Line 6". 

We further study the solid-state phase transition process from FE-ZB' to ML-$\beta$ phase with the same initial cell. As discussed in Fig. \ref{ZB_beta_WZ}, the transition process involves the step-wise migration of all three layers of O atoms. Here we followed the same moving order in which each line of the central-, bottom- and top-layer O atoms moves sequentially as shown in the ESI Movie S4. The final state is shown in Fig. \ref{MLdomain1}c and the corresponding energy evolution in Fig. \ref{MLdomain1}e. Eventually the whole cell is transformed into a 6$\times$6 ML-$\beta$ supercell in the final state, as the unit cell of the ML-$\beta$ phase consists of twice the number of atoms which compose the unit cell of the FE-ZB' phase. The nucleation of the initial ML-$\beta$ domain reduces the energy by 0.134 eV. After formation of the initial ML-$\beta$ domain, the following ML-$\beta$ domains grow with increasing reduction in energy, indicating an accelerating propagation of the domain transition. The final energy difference is 0.432 eV/unit cell which is very close the DFT data in Fig. \ref{ZB_beta_WZ}. We note that unlike the polarization reversal transition in Fig. \ref{MLdomain1}d, a single-atom jump frequently does not lead to stable configuration, as no saddle point is located on the energy pathway. Therefore, concerted transitions of lines of O atoms dominate the process.

\section{Discussion and Outlook}

Unlike the other III$_{2}$IV$_{3}$ compounds such as In$_{2}$Se$_{3}$, the bonding in Ga$_{2}$O$_{3}$ is mainly ionic in nature \cite{he2006electronic, Barman2019Mechanism}. The Bader charge analysis \cite{henkelman2006a, yu2011accurate} and the electron localization function \cite{ELF} on the (meta-) stable and saddle configurations indicate a similar ionicity of 2D Ga$_{2}$O$_{3}$ phases compared to bulk $\beta$-Ga$_{2}$O$_{3}$, as shown in the ESI Fig. S5. The energy difference between the FE and ML-$\beta$ phases is 0.430 eV/unit cell. The largest energy difference among the five metastable phases is $\sim$ 0.14 eV/unit cell. Interestingly, the metastable V2 phase has lower energy comparing to FE-ZB' and FE-WZ' phase ($\sim$ 0.07 eV/unit cell). The highest transition barrier predicted by our calculations is merely 0.311 eV per unit cell. With the underestimated margin due to the delocalization error of the approximated PBE functional as mentioned in section IIA, the transition barriers should be well below 1 eV per unit cell. The recent experimental studies on the Ga$_{2}$O$_{3}$ nanolayer indicate that amorphous thin films synthesized from liquid metal are stable under annealing at 600 K \cite{zavabeti2017liquid, wurdack2021ultrathin}. Our theoretical finding agrees with these experiments: with relatively small energy differences, comparable to the average atomic kinetic energy at room temperature (0.0388 eV/atom at 300 K), it is highly likely that several metastable phases co-exist. This may trap the system in a disordered amorphous-like structure, since the spontaneous growth of the crystalline phase requires crystal nuclei with high symmetry to grow beyond the critical size. The kinetics of the homogeneous and heterogeneous nucleation of the high symmetry phase is of great interest for future study. 

\section{Conclusions}

In summary, we systemically studied the phase transitions between three 2D low-energy configurations, namely FE-ZB', FE-WZ' and ML-$\beta$ of Ga$_{2}$O$_{3}$ monolayer using DFT and NEB calculations. The polarization-reversal transitions of the FE-ZB' phase can happen via direct jump or effective concerted movements, which can be further tuned by moderate external electric fields. The FE-WZ' phase exhibits higher transition barriers transforming to FE-ZB' phase. Multi-step solid-state transition paths from the metastable FE-ZB' and FE-WZ' phase to ML-$\beta$ phase are revealed as combinations of anisotropic lattice transformation and internal atomic migration. We further developed a machine-learning GAP potential to study the nucleation and growth of polarization reversed FE-ZB' domains and the FE-ZB'$\rightarrow$ML-$\beta$ transition. Different mechanisms of the domain wall propagation were identified. Based on our results, we expect that such 2D layers can be created experimentally either by mechanical exfoliation or epitaxial growth. In a broad perspective, this study adds important insight to the experimental synthesis of different 2D Ga$_{2}$O$_{3}$ structures. 

\begin{acknowledgments}
This work was supported by the National Natural Science Foundation of China (Grant 61904078) and the High-level University Fund (G02236002 and G02236005). 
This work was supported by the Center for Computational Science and Engineering at the Southern University of Science and Technology.
The authors are also grateful to the grants of computer power from CSC-IT, Center for Science, Finland. 
The authors would like to thank Haowen Mo at SUSTech for the helpful information on mechanical exfoliation of Ga$_{2}$O$_{3}$ thin film and Xinyu Wang at SUSTech for additional DFT data of bilayer Ga$_{2}$O$_{3}$ structures.
\end{acknowledgments}

\bibliographystyle{apsrev4-2}
\bibliography{new.bib}

\begin{thebibliography}{57}%
\makeatletter
\providecommand \@ifxundefined [1]{%
 \@ifx{#1\undefined}
}%
\providecommand \@ifnum [1]{%
 \ifnum #1\expandafter \@firstoftwo
 \else \expandafter \@secondoftwo
 \fi
}%
\providecommand \@ifx [1]{%
 \ifx #1\expandafter \@firstoftwo
 \else \expandafter \@secondoftwo
 \fi
}%
\providecommand \natexlab [1]{#1}%
\providecommand \enquote  [1]{``#1''}%
\providecommand \bibnamefont  [1]{#1}%
\providecommand \bibfnamefont [1]{#1}%
\providecommand \citenamefont [1]{#1}%
\providecommand \href@noop [0]{\@secondoftwo}%
\providecommand \href [0]{\begingroup \@sanitize@url \@href}%
\providecommand \@href[1]{\@@startlink{#1}\@@href}%
\providecommand \@@href[1]{\endgroup#1\@@endlink}%
\providecommand \@sanitize@url [0]{\catcode `\\12\catcode `\$12\catcode
  `\&12\catcode `\#12\catcode `\^12\catcode `\_12\catcode `\%12\relax}%
\providecommand \@@startlink[1]{}%
\providecommand \@@endlink[0]{}%
\providecommand \url  [0]{\begingroup\@sanitize@url \@url }%
\providecommand \@url [1]{\endgroup\@href {#1}{\urlprefix }}%
\providecommand \urlprefix  [0]{URL }%
\providecommand \Eprint [0]{\href }%
\providecommand \doibase [0]{https://doi.org/}%
\providecommand \selectlanguage [0]{\@gobble}%
\providecommand \bibinfo  [0]{\@secondoftwo}%
\providecommand \bibfield  [0]{\@secondoftwo}%
\providecommand \translation [1]{[#1]}%
\providecommand \BibitemOpen [0]{}%
\providecommand \bibitemStop [0]{}%
\providecommand \bibitemNoStop [0]{.\EOS\space}%
\providecommand \EOS [0]{\spacefactor3000\relax}%
\providecommand \BibitemShut  [1]{\csname bibitem#1\endcsname}%
\let\auto@bib@innerbib\@empty
\bibitem [{\citenamefont {{Pearton}}\ \emph {et~al.}(2018)\citenamefont
  {{Pearton}}, \citenamefont {{Yang}}, \citenamefont {{Cary}}, \citenamefont
  {{Ren}}, \citenamefont {{Kim}}, \citenamefont {{Tadjer}},\ and\ \citenamefont
  {{Mastro}}}]{pearton2018a}%
  \BibitemOpen
  \bibfield  {author} {\bibinfo {author} {\bibfnamefont {S.~J.}\ \bibnamefont
  {{Pearton}}}, \bibinfo {author} {\bibfnamefont {J.}~\bibnamefont {{Yang}}},
  \bibinfo {author} {\bibfnamefont {P.~H.}\ \bibnamefont {{Cary}}}, \bibinfo
  {author} {\bibfnamefont {F.}~\bibnamefont {{Ren}}}, \bibinfo {author}
  {\bibfnamefont {J.~H.}\ \bibnamefont {{Kim}}}, \bibinfo {author}
  {\bibfnamefont {M.~J.}\ \bibnamefont {{Tadjer}}},\ and\ \bibinfo {author}
  {\bibfnamefont {M.~A.}\ \bibnamefont {{Mastro}}},\ }\href@noop {} {\bibfield
  {journal} {\bibinfo  {journal} {Appl. Phys. Rev.}\ }\textbf {\bibinfo
  {volume} {5}},\ \bibinfo {pages} {11301} (\bibinfo {year}
  {2018})}\BibitemShut {NoStop}%
\bibitem [{\citenamefont {{Tsao}}\ \emph {et~al.}(2018)\citenamefont {{Tsao}},
  \citenamefont {{Chowdhury}}, \citenamefont {{Hollis}}, \citenamefont
  {{Jena}}, \citenamefont {{Johnson}}, \citenamefont {{Jones}}, \citenamefont
  {{Kaplar}}, \citenamefont {{Rajan}}, \citenamefont {de~{Walle}},
  \citenamefont {{Bellotti}}, \citenamefont {{Chua}}, \citenamefont
  {{Collazo}}, \citenamefont {{Coltrin}}, \citenamefont {{Cooper}},
  \citenamefont {{Evans}}, \citenamefont {{Graham}}, \citenamefont
  {{Grotjohn}}, \citenamefont {{Heller}}, \citenamefont {{Higashiwaki}},
  \citenamefont {{Islam}}, \citenamefont {{Juodawlkis}}, \citenamefont
  {{Khan}}, \citenamefont {{Koehler}}, \citenamefont {{Leach}}, \citenamefont
  {{Mishra}}, \citenamefont {{Nemanich}}, \citenamefont {{Pilawa-Podgurski}},
  \citenamefont {{Shealy}}, \citenamefont {{Sitar}}, \citenamefont {{Tadjer}},
  \citenamefont {{Witulski}}, \citenamefont {{Wraback}},\ and\ \citenamefont
  {{Simmons}}}]{tsao2018ultrawide}%
  \BibitemOpen
  \bibfield  {author} {\bibinfo {author} {\bibfnamefont {J.~Y.}\ \bibnamefont
  {{Tsao}}}, \bibinfo {author} {\bibfnamefont {S.}~\bibnamefont {{Chowdhury}}},
  \bibinfo {author} {\bibfnamefont {M.~A.}\ \bibnamefont {{Hollis}}}, \bibinfo
  {author} {\bibfnamefont {D.}~\bibnamefont {{Jena}}}, \bibinfo {author}
  {\bibfnamefont {N.~M.}\ \bibnamefont {{Johnson}}}, \bibinfo {author}
  {\bibfnamefont {K.~A.}\ \bibnamefont {{Jones}}}, \bibinfo {author}
  {\bibfnamefont {R.~J.}\ \bibnamefont {{Kaplar}}}, \bibinfo {author}
  {\bibfnamefont {S.}~\bibnamefont {{Rajan}}}, \bibinfo {author} {\bibfnamefont
  {C.~G.~V.}\ \bibnamefont {de~{Walle}}}, \bibinfo {author} {\bibfnamefont
  {E.}~\bibnamefont {{Bellotti}}}, \bibinfo {author} {\bibfnamefont {C.~L.}\
  \bibnamefont {{Chua}}}, \bibinfo {author} {\bibfnamefont {R.}~\bibnamefont
  {{Collazo}}}, \bibinfo {author} {\bibfnamefont {M.~E.}\ \bibnamefont
  {{Coltrin}}}, \bibinfo {author} {\bibfnamefont {J.~A.}\ \bibnamefont
  {{Cooper}}}, \bibinfo {author} {\bibfnamefont {K.~R.}\ \bibnamefont
  {{Evans}}}, \bibinfo {author} {\bibfnamefont {S.}~\bibnamefont {{Graham}}},
  \bibinfo {author} {\bibfnamefont {T.~A.}\ \bibnamefont {{Grotjohn}}},
  \bibinfo {author} {\bibfnamefont {E.~R.}\ \bibnamefont {{Heller}}}, \bibinfo
  {author} {\bibfnamefont {M.}~\bibnamefont {{Higashiwaki}}}, \bibinfo {author}
  {\bibfnamefont {M.~S.}\ \bibnamefont {{Islam}}}, \bibinfo {author}
  {\bibfnamefont {P.~W.}\ \bibnamefont {{Juodawlkis}}}, \bibinfo {author}
  {\bibfnamefont {M.~A.}\ \bibnamefont {{Khan}}}, \bibinfo {author}
  {\bibfnamefont {A.~D.}\ \bibnamefont {{Koehler}}}, \bibinfo {author}
  {\bibfnamefont {J.~H.}\ \bibnamefont {{Leach}}}, \bibinfo {author}
  {\bibfnamefont {U.~K.}\ \bibnamefont {{Mishra}}}, \bibinfo {author}
  {\bibfnamefont {R.}~\bibnamefont {{Nemanich}}}, \bibinfo {author}
  {\bibfnamefont {R.~C.}\ \bibnamefont {{Pilawa-Podgurski}}}, \bibinfo {author}
  {\bibfnamefont {J.~B.}\ \bibnamefont {{Shealy}}}, \bibinfo {author}
  {\bibfnamefont {Z.}~\bibnamefont {{Sitar}}}, \bibinfo {author} {\bibfnamefont
  {M.~J.}\ \bibnamefont {{Tadjer}}}, \bibinfo {author} {\bibfnamefont {A.~F.}\
  \bibnamefont {{Witulski}}}, \bibinfo {author} {\bibfnamefont
  {M.}~\bibnamefont {{Wraback}}},\ and\ \bibinfo {author} {\bibfnamefont
  {J.~A.}\ \bibnamefont {{Simmons}}},\ }\href@noop {} {\bibfield  {journal}
  {\bibinfo  {journal} {Adv. Electron. Mater.}\ }\textbf {\bibinfo {volume}
  {4}},\ \bibinfo {pages} {1600501} (\bibinfo {year} {2018})}\BibitemShut
  {NoStop}%
\bibitem [{\citenamefont {Orita}\ \emph {et~al.}(2000)\citenamefont {Orita},
  \citenamefont {Ohta}, \citenamefont {Hirano},\ and\ \citenamefont
  {Hosono}}]{Orita2000Deep}%
  \BibitemOpen
  \bibfield  {author} {\bibinfo {author} {\bibfnamefont {M.}~\bibnamefont
  {Orita}}, \bibinfo {author} {\bibfnamefont {H.}~\bibnamefont {Ohta}},
  \bibinfo {author} {\bibfnamefont {M.}~\bibnamefont {Hirano}},\ and\ \bibinfo
  {author} {\bibfnamefont {H.}~\bibnamefont {Hosono}},\ }\href
  {https://doi.org/10.1063/1.1330559} {\bibfield  {journal} {\bibinfo
  {journal} {Appl. Phys. Lett.}\ }\textbf {\bibinfo {volume} {77}},\ \bibinfo
  {pages} {4166} (\bibinfo {year} {2000})}\BibitemShut {NoStop}%
\bibitem [{\citenamefont {Peelaers}\ and\ \citenamefont {Van~de
  Walle}(2015)}]{Peelaers2015Bri}%
  \BibitemOpen
  \bibfield  {author} {\bibinfo {author} {\bibfnamefont {H.}~\bibnamefont
  {Peelaers}}\ and\ \bibinfo {author} {\bibfnamefont {C.~G.}\ \bibnamefont
  {Van~de Walle}},\ }\href
  {https://doi.org/https://doi.org/10.1002/pssb.201451551} {\bibfield
  {journal} {\bibinfo  {journal} {Phys. Status Solidi B}\ }\textbf {\bibinfo
  {volume} {252}},\ \bibinfo {pages} {828} (\bibinfo {year}
  {2015})}\BibitemShut {NoStop}%
\bibitem [{\citenamefont {Furthm\"uller}\ and\ \citenamefont
  {Bechstedt}(2016)}]{Furthm2016Quasi}%
  \BibitemOpen
  \bibfield  {author} {\bibinfo {author} {\bibfnamefont {J.}~\bibnamefont
  {Furthm\"uller}}\ and\ \bibinfo {author} {\bibfnamefont {F.}~\bibnamefont
  {Bechstedt}},\ }\href {https://doi.org/10.1103/PhysRevB.93.115204} {\bibfield
   {journal} {\bibinfo  {journal} {Phys. Rev. B}\ }\textbf {\bibinfo {volume}
  {93}},\ \bibinfo {pages} {115204} (\bibinfo {year} {2016})}\BibitemShut
  {NoStop}%
\bibitem [{\citenamefont {{Wurdack}}\ \emph {et~al.}(2021)\citenamefont
  {{Wurdack}}, \citenamefont {{Yun}}, \citenamefont {{Estrecho}}, \citenamefont
  {{Syed}}, \citenamefont {{Bhattacharyya}}, \citenamefont {{Pieczarka}},
  \citenamefont {{Zavabeti}}, \citenamefont {{Chen}}, \citenamefont {{Haas}},
  \citenamefont {{Müller}}, \citenamefont {{Lockrey}}, \citenamefont {{Bao}},
  \citenamefont {{Schneider}}, \citenamefont {{Lu}}, \citenamefont {{Fuhrer}},
  \citenamefont {{Truscott}}, \citenamefont {{Daeneke}},\ and\ \citenamefont
  {{Ostrovskaya}}}]{wurdack2021ultrathin}%
  \BibitemOpen
  \bibfield  {author} {\bibinfo {author} {\bibfnamefont {M.}~\bibnamefont
  {{Wurdack}}}, \bibinfo {author} {\bibfnamefont {T.}~\bibnamefont {{Yun}}},
  \bibinfo {author} {\bibfnamefont {E.}~\bibnamefont {{Estrecho}}}, \bibinfo
  {author} {\bibfnamefont {N.}~\bibnamefont {{Syed}}}, \bibinfo {author}
  {\bibfnamefont {S.}~\bibnamefont {{Bhattacharyya}}}, \bibinfo {author}
  {\bibfnamefont {M.}~\bibnamefont {{Pieczarka}}}, \bibinfo {author}
  {\bibfnamefont {A.}~\bibnamefont {{Zavabeti}}}, \bibinfo {author}
  {\bibfnamefont {S.}~\bibnamefont {{Chen}}}, \bibinfo {author} {\bibfnamefont
  {B.}~\bibnamefont {{Haas}}}, \bibinfo {author} {\bibfnamefont
  {J.}~\bibnamefont {{Müller}}}, \bibinfo {author} {\bibfnamefont {M.~N.}\
  \bibnamefont {{Lockrey}}}, \bibinfo {author} {\bibfnamefont {Q.}~\bibnamefont
  {{Bao}}}, \bibinfo {author} {\bibfnamefont {C.}~\bibnamefont {{Schneider}}},
  \bibinfo {author} {\bibfnamefont {Y.}~\bibnamefont {{Lu}}}, \bibinfo {author}
  {\bibfnamefont {M.~S.}\ \bibnamefont {{Fuhrer}}}, \bibinfo {author}
  {\bibfnamefont {A.~G.}\ \bibnamefont {{Truscott}}}, \bibinfo {author}
  {\bibfnamefont {T.}~\bibnamefont {{Daeneke}}},\ and\ \bibinfo {author}
  {\bibfnamefont {E.~A.}\ \bibnamefont {{Ostrovskaya}}},\ }\href@noop {}
  {\bibfield  {journal} {\bibinfo  {journal} {Adv. Mater.}\ }\textbf {\bibinfo
  {volume} {33}},\ \bibinfo {pages} {2005732} (\bibinfo {year}
  {2021})}\BibitemShut {NoStop}%
\bibitem [{\citenamefont {{Wang}}\ \emph {et~al.}(2021)\citenamefont {{Wang}},
  \citenamefont {{Zhang}}, \citenamefont {{Xu}}, \citenamefont {{Zhang}},
  \citenamefont {{Feng}}, \citenamefont {{Zhang}}, \citenamefont {{Ning}},
  \citenamefont {{Zhao}}, \citenamefont {{Zhou}},\ and\ \citenamefont
  {{Hao}}}]{wang2021progresses}%
  \BibitemOpen
  \bibfield  {author} {\bibinfo {author} {\bibfnamefont {C.}~\bibnamefont
  {{Wang}}}, \bibinfo {author} {\bibfnamefont {J.}~\bibnamefont {{Zhang}}},
  \bibinfo {author} {\bibfnamefont {S.}~\bibnamefont {{Xu}}}, \bibinfo {author}
  {\bibfnamefont {C.}~\bibnamefont {{Zhang}}}, \bibinfo {author} {\bibfnamefont
  {Q.}~\bibnamefont {{Feng}}}, \bibinfo {author} {\bibfnamefont
  {Y.}~\bibnamefont {{Zhang}}}, \bibinfo {author} {\bibfnamefont
  {J.}~\bibnamefont {{Ning}}}, \bibinfo {author} {\bibfnamefont
  {S.}~\bibnamefont {{Zhao}}}, \bibinfo {author} {\bibfnamefont
  {H.}~\bibnamefont {{Zhou}}},\ and\ \bibinfo {author} {\bibfnamefont
  {Y.}~\bibnamefont {{Hao}}},\ }\bibfield  {journal} {\bibinfo  {journal} {J.
  Phys. D: Appl. Phys.}\ }\href {https://doi.org/10.1088/1361-6463/abe158}
  {10.1088/1361-6463/abe158} (\bibinfo {year} {2021})\BibitemShut {NoStop}%
\bibitem [{\citenamefont {Peelaers}\ and\ \citenamefont {Van~de
  Walle}(2017)}]{peelaers2017lack}%
  \BibitemOpen
  \bibfield  {author} {\bibinfo {author} {\bibfnamefont {H.}~\bibnamefont
  {Peelaers}}\ and\ \bibinfo {author} {\bibfnamefont {C.~G.}\ \bibnamefont
  {Van~de Walle}},\ }\href {https://doi.org/10.1103/PhysRevB.96.081409}
  {\bibfield  {journal} {\bibinfo  {journal} {Phys. Rev. B}\ }\textbf {\bibinfo
  {volume} {96}},\ \bibinfo {pages} {081409(R)} (\bibinfo {year}
  {2017})}\BibitemShut {NoStop}%
\bibitem [{\citenamefont {Zhou}\ \emph {et~al.}(2017)\citenamefont {Zhou},
  \citenamefont {Maize}, \citenamefont {Qiu}, \citenamefont {Shakouri},\ and\
  \citenamefont {Ye}}]{Zhou2017insulator}%
  \BibitemOpen
  \bibfield  {author} {\bibinfo {author} {\bibfnamefont {H.}~\bibnamefont
  {Zhou}}, \bibinfo {author} {\bibfnamefont {K.}~\bibnamefont {Maize}},
  \bibinfo {author} {\bibfnamefont {G.}~\bibnamefont {Qiu}}, \bibinfo {author}
  {\bibfnamefont {A.}~\bibnamefont {Shakouri}},\ and\ \bibinfo {author}
  {\bibfnamefont {P.~D.}\ \bibnamefont {Ye}},\ }\href
  {https://doi.org/10.1063/1.5000735} {\bibfield  {journal} {\bibinfo
  {journal} {Appl. Phys. Lett.}\ }\textbf {\bibinfo {volume} {111}},\ \bibinfo
  {pages} {092102} (\bibinfo {year} {2017})}\BibitemShut {NoStop}%
\bibitem [{\citenamefont {{Kwon}}\ \emph {et~al.}(2017)\citenamefont {{Kwon}},
  \citenamefont {{Lee}}, \citenamefont {{Oh}}, \citenamefont {{Kim}},
  \citenamefont {{Pearton}},\ and\ \citenamefont {{Ren}}}]{kwon2017tuning}%
  \BibitemOpen
  \bibfield  {author} {\bibinfo {author} {\bibfnamefont {Y.}~\bibnamefont
  {{Kwon}}}, \bibinfo {author} {\bibfnamefont {G.}~\bibnamefont {{Lee}}},
  \bibinfo {author} {\bibfnamefont {S.}~\bibnamefont {{Oh}}}, \bibinfo {author}
  {\bibfnamefont {J.~H.}\ \bibnamefont {{Kim}}}, \bibinfo {author}
  {\bibfnamefont {S.~J.}\ \bibnamefont {{Pearton}}},\ and\ \bibinfo {author}
  {\bibfnamefont {F.}~\bibnamefont {{Ren}}},\ }\href@noop {} {\bibfield
  {journal} {\bibinfo  {journal} {Appl. Phys. Lett.}\ }\textbf {\bibinfo
  {volume} {110}},\ \bibinfo {pages} {131901} (\bibinfo {year}
  {2017})}\BibitemShut {NoStop}%
\bibitem [{\citenamefont {Barman}\ and\ \citenamefont
  {Huda}(2019)}]{Barman2019Mechanism}%
  \BibitemOpen
  \bibfield  {author} {\bibinfo {author} {\bibfnamefont {S.~K.}\ \bibnamefont
  {Barman}}\ and\ \bibinfo {author} {\bibfnamefont {M.~N.}\ \bibnamefont
  {Huda}},\ }\href {https://doi.org/https://doi.org/10.1002/pssr.201800554}
  {\bibfield  {journal} {\bibinfo  {journal} {Phys. Status Solidi RRL}\
  }\textbf {\bibinfo {volume} {13}},\ \bibinfo {pages} {1800554} (\bibinfo
  {year} {2019})}\BibitemShut {NoStop}%
\bibitem [{\citenamefont {Hwang}\ \emph {et~al.}(2014)\citenamefont {Hwang},
  \citenamefont {Verma}, \citenamefont {Peelaers}, \citenamefont {Protasenko},
  \citenamefont {Rouvimov}, \citenamefont {(Grace)~Xing}, \citenamefont
  {Seabaugh}, \citenamefont {Haensch}, \citenamefont {de~Walle}, \citenamefont
  {Galazka}, \citenamefont {Albrecht}, \citenamefont {Fornari},\ and\
  \citenamefont {Jena}}]{Hwang2014High}%
  \BibitemOpen
  \bibfield  {author} {\bibinfo {author} {\bibfnamefont {W.~S.}\ \bibnamefont
  {Hwang}}, \bibinfo {author} {\bibfnamefont {A.}~\bibnamefont {Verma}},
  \bibinfo {author} {\bibfnamefont {H.}~\bibnamefont {Peelaers}}, \bibinfo
  {author} {\bibfnamefont {V.}~\bibnamefont {Protasenko}}, \bibinfo {author}
  {\bibfnamefont {S.}~\bibnamefont {Rouvimov}}, \bibinfo {author}
  {\bibfnamefont {H.}~\bibnamefont {(Grace)~Xing}}, \bibinfo {author}
  {\bibfnamefont {A.}~\bibnamefont {Seabaugh}}, \bibinfo {author}
  {\bibfnamefont {W.}~\bibnamefont {Haensch}}, \bibinfo {author} {\bibfnamefont
  {C.~V.}\ \bibnamefont {de~Walle}}, \bibinfo {author} {\bibfnamefont
  {Z.}~\bibnamefont {Galazka}}, \bibinfo {author} {\bibfnamefont
  {M.}~\bibnamefont {Albrecht}}, \bibinfo {author} {\bibfnamefont
  {R.}~\bibnamefont {Fornari}},\ and\ \bibinfo {author} {\bibfnamefont
  {D.}~\bibnamefont {Jena}},\ }\href {https://doi.org/10.1063/1.4879800}
  {\bibfield  {journal} {\bibinfo  {journal} {Appl. Phys. Lett.}\ }\textbf
  {\bibinfo {volume} {104}},\ \bibinfo {pages} {203111} (\bibinfo {year}
  {2014})}\BibitemShut {NoStop}%
\bibitem [{\citenamefont {{Zhou}}\ \emph {et~al.}(2017)\citenamefont {{Zhou}},
  \citenamefont {{Si}}, \citenamefont {{Alghamdi}}, \citenamefont {{Qiu}},
  \citenamefont {{Yang}},\ and\ \citenamefont {{Ye}}}]{zhou2017high}%
  \BibitemOpen
  \bibfield  {author} {\bibinfo {author} {\bibfnamefont {H.}~\bibnamefont
  {{Zhou}}}, \bibinfo {author} {\bibfnamefont {M.}~\bibnamefont {{Si}}},
  \bibinfo {author} {\bibfnamefont {S.}~\bibnamefont {{Alghamdi}}}, \bibinfo
  {author} {\bibfnamefont {G.}~\bibnamefont {{Qiu}}}, \bibinfo {author}
  {\bibfnamefont {L.}~\bibnamefont {{Yang}}},\ and\ \bibinfo {author}
  {\bibfnamefont {P.~D.}\ \bibnamefont {{Ye}}},\ }\href@noop {} {\bibfield
  {journal} {\bibinfo  {journal} {IEEE Electron. Device Lett.}\ }\textbf
  {\bibinfo {volume} {38}},\ \bibinfo {pages} {103} (\bibinfo {year}
  {2017})}\BibitemShut {NoStop}%
\bibitem [{\citenamefont {Chen}\ \emph {et~al.}(2019)\citenamefont {Chen},
  \citenamefont {Li}, \citenamefont {Ma}, \citenamefont {Huang}, \citenamefont
  {Ji}, \citenamefont {Xia}, \citenamefont {Lu},\ and\ \citenamefont
  {Zhang}}]{Chen2019Ohmic}%
  \BibitemOpen
  \bibfield  {author} {\bibinfo {author} {\bibfnamefont {J.-X.}\ \bibnamefont
  {Chen}}, \bibinfo {author} {\bibfnamefont {X.-X.}\ \bibnamefont {Li}},
  \bibinfo {author} {\bibfnamefont {H.-P.}\ \bibnamefont {Ma}}, \bibinfo
  {author} {\bibfnamefont {W.}~\bibnamefont {Huang}}, \bibinfo {author}
  {\bibfnamefont {Z.-G.}\ \bibnamefont {Ji}}, \bibinfo {author} {\bibfnamefont
  {C.}~\bibnamefont {Xia}}, \bibinfo {author} {\bibfnamefont {H.-L.}\
  \bibnamefont {Lu}},\ and\ \bibinfo {author} {\bibfnamefont {D.~W.}\
  \bibnamefont {Zhang}},\ }\href {https://doi.org/10.1021/acsami.9b09166}
  {\bibfield  {journal} {\bibinfo  {journal} {ACS Appl. Mater. Inter.}\
  }\textbf {\bibinfo {volume} {11}},\ \bibinfo {pages} {32127} (\bibinfo {year}
  {2019})}\BibitemShut {NoStop}%
\bibitem [{\citenamefont {{Chandiran}}\ \emph {et~al.}(2012)\citenamefont
  {{Chandiran}}, \citenamefont {{Tetreault}}, \citenamefont {{Humphry-Baker}},
  \citenamefont {{Kessler}}, \citenamefont {{Baranoff}}, \citenamefont {{Yi}},
  \citenamefont {{Nazeeruddin}},\ and\ \citenamefont
  {{Grätzel}}}]{chandiran2012subnanometer}%
  \BibitemOpen
  \bibfield  {author} {\bibinfo {author} {\bibfnamefont {A.~K.}\ \bibnamefont
  {{Chandiran}}}, \bibinfo {author} {\bibfnamefont {N.}~\bibnamefont
  {{Tetreault}}}, \bibinfo {author} {\bibfnamefont {R.}~\bibnamefont
  {{Humphry-Baker}}}, \bibinfo {author} {\bibfnamefont {F.}~\bibnamefont
  {{Kessler}}}, \bibinfo {author} {\bibfnamefont {E.}~\bibnamefont
  {{Baranoff}}}, \bibinfo {author} {\bibfnamefont {C.}~\bibnamefont {{Yi}}},
  \bibinfo {author} {\bibfnamefont {M.~K.}\ \bibnamefont {{Nazeeruddin}}},\
  and\ \bibinfo {author} {\bibfnamefont {M.}~\bibnamefont {{Grätzel}}},\
  }\href@noop {} {\bibfield  {journal} {\bibinfo  {journal} {Nano Lett.}\
  }\textbf {\bibinfo {volume} {12}},\ \bibinfo {pages} {3941} (\bibinfo {year}
  {2012})}\BibitemShut {NoStop}%
\bibitem [{\citenamefont {Zhang}\ \emph {et~al.}(2020)\citenamefont {Zhang},
  \citenamefont {Li}, \citenamefont {Zhang}, \citenamefont {Feng},
  \citenamefont {Ning}, \citenamefont {Zhang}, \citenamefont {Zhang},\ and\
  \citenamefont {Hao}}]{ZHANG2020157810}%
  \BibitemOpen
  \bibfield  {author} {\bibinfo {author} {\bibfnamefont {T.}~\bibnamefont
  {Zhang}}, \bibinfo {author} {\bibfnamefont {Y.}~\bibnamefont {Li}}, \bibinfo
  {author} {\bibfnamefont {Y.}~\bibnamefont {Zhang}}, \bibinfo {author}
  {\bibfnamefont {Q.}~\bibnamefont {Feng}}, \bibinfo {author} {\bibfnamefont
  {J.}~\bibnamefont {Ning}}, \bibinfo {author} {\bibfnamefont {C.}~\bibnamefont
  {Zhang}}, \bibinfo {author} {\bibfnamefont {J.}~\bibnamefont {Zhang}},\ and\
  \bibinfo {author} {\bibfnamefont {Y.}~\bibnamefont {Hao}},\ }\href
  {https://doi.org/https://doi.org/10.1016/j.jallcom.2020.157810} {\bibfield
  {journal} {\bibinfo  {journal} {J. Alloys Compd.}\ ,\ \bibinfo {pages}
  {157810}} (\bibinfo {year} {2020})}\BibitemShut {NoStop}%
\bibitem [{\citenamefont {Zavabeti}\ \emph {et~al.}(2017)\citenamefont
  {Zavabeti}, \citenamefont {Ou}, \citenamefont {Carey}, \citenamefont {Syed},
  \citenamefont {Orrell-Trigg}, \citenamefont {Mayes}, \citenamefont {Xu},
  \citenamefont {Kavehei}, \citenamefont {O’Mullane}, \citenamefont {Kaner}
  \emph {et~al.}}]{zavabeti2017liquid}%
  \BibitemOpen
  \bibfield  {author} {\bibinfo {author} {\bibfnamefont {A.}~\bibnamefont
  {Zavabeti}}, \bibinfo {author} {\bibfnamefont {J.~Z.}\ \bibnamefont {Ou}},
  \bibinfo {author} {\bibfnamefont {B.~J.}\ \bibnamefont {Carey}}, \bibinfo
  {author} {\bibfnamefont {N.}~\bibnamefont {Syed}}, \bibinfo {author}
  {\bibfnamefont {R.}~\bibnamefont {Orrell-Trigg}}, \bibinfo {author}
  {\bibfnamefont {E.~L.}\ \bibnamefont {Mayes}}, \bibinfo {author}
  {\bibfnamefont {C.}~\bibnamefont {Xu}}, \bibinfo {author} {\bibfnamefont
  {O.}~\bibnamefont {Kavehei}}, \bibinfo {author} {\bibfnamefont {A.~P.}\
  \bibnamefont {O’Mullane}}, \bibinfo {author} {\bibfnamefont {R.~B.}\
  \bibnamefont {Kaner}}, \emph {et~al.},\ }\href@noop {} {\bibfield  {journal}
  {\bibinfo  {journal} {Science}\ }\textbf {\bibinfo {volume} {358}},\ \bibinfo
  {pages} {332} (\bibinfo {year} {2017})}\BibitemShut {NoStop}%
\bibitem [{\citenamefont {{Wang}}\ \emph {et~al.}(2019)\citenamefont {{Wang}},
  \citenamefont {{Xu}}, \citenamefont {{Zhang}}, \citenamefont {{Qiao}},
  \citenamefont {{Wu}}, \citenamefont {{Wang}}, \citenamefont {{Zhang}},
  \citenamefont {{Liang}}, \citenamefont {{Zhang}}, \citenamefont {{Zhang}},
  \citenamefont {{Chen}}, \citenamefont {{Xie}}, \citenamefont {{Zong}},
  \citenamefont {{Shan}}, \citenamefont {{Guo}}, \citenamefont {{Willinger}},
  \citenamefont {{Wu}}, \citenamefont {{Li}}, \citenamefont {{Wang}},
  \citenamefont {{Gao}}, \citenamefont {{Wu}}, \citenamefont {{Zhang}},
  \citenamefont {{Jiang}}, \citenamefont {{Yu}}, \citenamefont {{Wang}},
  \citenamefont {{Bai}}, \citenamefont {{Wang}}, \citenamefont {{Ding}},\ and\
  \citenamefont {{Liu}}}]{wang2019epitaxial}%
  \BibitemOpen
  \bibfield  {author} {\bibinfo {author} {\bibfnamefont {L.}~\bibnamefont
  {{Wang}}}, \bibinfo {author} {\bibfnamefont {X.}~\bibnamefont {{Xu}}},
  \bibinfo {author} {\bibfnamefont {L.}~\bibnamefont {{Zhang}}}, \bibinfo
  {author} {\bibfnamefont {R.}~\bibnamefont {{Qiao}}}, \bibinfo {author}
  {\bibfnamefont {M.}~\bibnamefont {{Wu}}}, \bibinfo {author} {\bibfnamefont
  {Z.}~\bibnamefont {{Wang}}}, \bibinfo {author} {\bibfnamefont
  {S.}~\bibnamefont {{Zhang}}}, \bibinfo {author} {\bibfnamefont
  {J.}~\bibnamefont {{Liang}}}, \bibinfo {author} {\bibfnamefont
  {Z.}~\bibnamefont {{Zhang}}}, \bibinfo {author} {\bibfnamefont
  {Z.}~\bibnamefont {{Zhang}}}, \bibinfo {author} {\bibfnamefont
  {W.}~\bibnamefont {{Chen}}}, \bibinfo {author} {\bibfnamefont
  {X.}~\bibnamefont {{Xie}}}, \bibinfo {author} {\bibfnamefont
  {J.}~\bibnamefont {{Zong}}}, \bibinfo {author} {\bibfnamefont
  {Y.}~\bibnamefont {{Shan}}}, \bibinfo {author} {\bibfnamefont
  {Y.}~\bibnamefont {{Guo}}}, \bibinfo {author} {\bibfnamefont {M.~G.}\
  \bibnamefont {{Willinger}}}, \bibinfo {author} {\bibfnamefont
  {H.}~\bibnamefont {{Wu}}}, \bibinfo {author} {\bibfnamefont {Q.}~\bibnamefont
  {{Li}}}, \bibinfo {author} {\bibfnamefont {W.}~\bibnamefont {{Wang}}},
  \bibinfo {author} {\bibfnamefont {P.}~\bibnamefont {{Gao}}}, \bibinfo
  {author} {\bibfnamefont {S.}~\bibnamefont {{Wu}}}, \bibinfo {author}
  {\bibfnamefont {Y.}~\bibnamefont {{Zhang}}}, \bibinfo {author} {\bibfnamefont
  {Y.}~\bibnamefont {{Jiang}}}, \bibinfo {author} {\bibfnamefont
  {D.}~\bibnamefont {{Yu}}}, \bibinfo {author} {\bibfnamefont {E.}~\bibnamefont
  {{Wang}}}, \bibinfo {author} {\bibfnamefont {X.}~\bibnamefont {{Bai}}},
  \bibinfo {author} {\bibfnamefont {Z.-J.}\ \bibnamefont {{Wang}}}, \bibinfo
  {author} {\bibfnamefont {F.}~\bibnamefont {{Ding}}},\ and\ \bibinfo {author}
  {\bibfnamefont {K.}~\bibnamefont {{Liu}}},\ }\href@noop {} {\bibfield
  {journal} {\bibinfo  {journal} {Nature}\ }\textbf {\bibinfo {volume} {570}},\
  \bibinfo {pages} {91} (\bibinfo {year} {2019})}\BibitemShut {NoStop}%
\bibitem [{\citenamefont {{Dong}}\ \emph {et~al.}(2020)\citenamefont {{Dong}},
  \citenamefont {{Zhang}}, \citenamefont {{Dai}},\ and\ \citenamefont
  {{Ding}}}]{dong2020the}%
  \BibitemOpen
  \bibfield  {author} {\bibinfo {author} {\bibfnamefont {J.}~\bibnamefont
  {{Dong}}}, \bibinfo {author} {\bibfnamefont {L.}~\bibnamefont {{Zhang}}},
  \bibinfo {author} {\bibfnamefont {X.}~\bibnamefont {{Dai}}},\ and\ \bibinfo
  {author} {\bibfnamefont {F.}~\bibnamefont {{Ding}}},\ }\href@noop {}
  {\bibfield  {journal} {\bibinfo  {journal} {Nat. Commun.}\ }\textbf {\bibinfo
  {volume} {11}},\ \bibinfo {pages} {5862} (\bibinfo {year}
  {2020})}\BibitemShut {NoStop}%
\bibitem [{\citenamefont {{Chen}}\ \emph {et~al.}(2020)\citenamefont {{Chen}},
  \citenamefont {{Chuu}}, \citenamefont {{Tseng}}, \citenamefont {{Wen}},
  \citenamefont {{Wong}}, \citenamefont {{Pan}}, \citenamefont {{Li}},
  \citenamefont {{Chao}}, \citenamefont {{Chueh}}, \citenamefont {{Zhang}},
  \citenamefont {{Fu}}, \citenamefont {{Yakobson}}, \citenamefont {{Chang}},\
  and\ \citenamefont {{Li}}}]{chen2020wafer}%
  \BibitemOpen
  \bibfield  {author} {\bibinfo {author} {\bibfnamefont {T.~A.}\ \bibnamefont
  {{Chen}}}, \bibinfo {author} {\bibfnamefont {C.~P.}\ \bibnamefont {{Chuu}}},
  \bibinfo {author} {\bibfnamefont {C.~C.}\ \bibnamefont {{Tseng}}}, \bibinfo
  {author} {\bibfnamefont {C.~K.}\ \bibnamefont {{Wen}}}, \bibinfo {author}
  {\bibfnamefont {H.~S.}\ \bibnamefont {{Wong}}}, \bibinfo {author}
  {\bibfnamefont {S.}~\bibnamefont {{Pan}}}, \bibinfo {author} {\bibfnamefont
  {R.}~\bibnamefont {{Li}}}, \bibinfo {author} {\bibfnamefont {T.~A.}\
  \bibnamefont {{Chao}}}, \bibinfo {author} {\bibfnamefont {W.~C.}\
  \bibnamefont {{Chueh}}}, \bibinfo {author} {\bibfnamefont {Y.}~\bibnamefont
  {{Zhang}}}, \bibinfo {author} {\bibfnamefont {Q.}~\bibnamefont {{Fu}}},
  \bibinfo {author} {\bibfnamefont {B.~I.}\ \bibnamefont {{Yakobson}}},
  \bibinfo {author} {\bibfnamefont {W.~H.}\ \bibnamefont {{Chang}}},\ and\
  \bibinfo {author} {\bibfnamefont {L.~J.}\ \bibnamefont {{Li}}},\ }\href@noop
  {} {\bibfield  {journal} {\bibinfo  {journal} {Nature}\ }\textbf {\bibinfo
  {volume} {579}},\ \bibinfo {pages} {219} (\bibinfo {year}
  {2020})}\BibitemShut {NoStop}%
\bibitem [{\citenamefont {{Zhang}}\ \emph {et~al.}(2021)\citenamefont
  {{Zhang}}, \citenamefont {{Xu}}, \citenamefont {{Yao}}, \citenamefont
  {{Jannat}}, \citenamefont {{Ren}}, \citenamefont {{Field}}, \citenamefont
  {{Wen}}, \citenamefont {{Zhou}}, \citenamefont {{Zavabeti}},\ and\
  \citenamefont {{Ou}}}]{zhang2021hexagonal}%
  \BibitemOpen
  \bibfield  {author} {\bibinfo {author} {\bibfnamefont {B.~Y.}\ \bibnamefont
  {{Zhang}}}, \bibinfo {author} {\bibfnamefont {K.}~\bibnamefont {{Xu}}},
  \bibinfo {author} {\bibfnamefont {Q.}~\bibnamefont {{Yao}}}, \bibinfo
  {author} {\bibfnamefont {A.}~\bibnamefont {{Jannat}}}, \bibinfo {author}
  {\bibfnamefont {G.}~\bibnamefont {{Ren}}}, \bibinfo {author} {\bibfnamefont
  {M.~R.}\ \bibnamefont {{Field}}}, \bibinfo {author} {\bibfnamefont
  {X.}~\bibnamefont {{Wen}}}, \bibinfo {author} {\bibfnamefont
  {C.}~\bibnamefont {{Zhou}}}, \bibinfo {author} {\bibfnamefont
  {A.}~\bibnamefont {{Zavabeti}}},\ and\ \bibinfo {author} {\bibfnamefont
  {J.~Z.}\ \bibnamefont {{Ou}}},\ }\href@noop {} {\bibfield  {journal}
  {\bibinfo  {journal} {Nat. Mater.}\ ,\ \bibinfo {pages} {1}} (\bibinfo {year}
  {2021})}\BibitemShut {NoStop}%
\bibitem [{\citenamefont {{Liao}}\ \emph {et~al.}(2020)\citenamefont {{Liao}},
  \citenamefont {{Zhang}}, \citenamefont {{Gao}}, \citenamefont {{Qian}},\ and\
  \citenamefont {{Hua}}}]{liao2020tunable}%
  \BibitemOpen
  \bibfield  {author} {\bibinfo {author} {\bibfnamefont {Y.}~\bibnamefont
  {{Liao}}}, \bibinfo {author} {\bibfnamefont {Z.}~\bibnamefont {{Zhang}}},
  \bibinfo {author} {\bibfnamefont {Z.}~\bibnamefont {{Gao}}}, \bibinfo
  {author} {\bibfnamefont {Q.}~\bibnamefont {{Qian}}},\ and\ \bibinfo {author}
  {\bibfnamefont {M.}~\bibnamefont {{Hua}}},\ }\href@noop {} {\bibfield
  {journal} {\bibinfo  {journal} {ACS Appl. Mater. Inter.}\ }\textbf {\bibinfo
  {volume} {12}},\ \bibinfo {pages} {30659} (\bibinfo {year}
  {2020})}\BibitemShut {NoStop}%
\bibitem [{\citenamefont {{Ding}}\ \emph {et~al.}(2017)\citenamefont {{Ding}},
  \citenamefont {{Zhu}}, \citenamefont {{Wang}}, \citenamefont {{Gao}},
  \citenamefont {{Xiao}}, \citenamefont {{Gu}}, \citenamefont {{Zhang}},\ and\
  \citenamefont {{Zhu}}}]{ding2017prediction}%
  \BibitemOpen
  \bibfield  {author} {\bibinfo {author} {\bibfnamefont {W.}~\bibnamefont
  {{Ding}}}, \bibinfo {author} {\bibfnamefont {J.}~\bibnamefont {{Zhu}}},
  \bibinfo {author} {\bibfnamefont {Z.}~\bibnamefont {{Wang}}}, \bibinfo
  {author} {\bibfnamefont {Y.}~\bibnamefont {{Gao}}}, \bibinfo {author}
  {\bibfnamefont {D.}~\bibnamefont {{Xiao}}}, \bibinfo {author} {\bibfnamefont
  {Y.}~\bibnamefont {{Gu}}}, \bibinfo {author} {\bibfnamefont {Z.}~\bibnamefont
  {{Zhang}}},\ and\ \bibinfo {author} {\bibfnamefont {W.}~\bibnamefont
  {{Zhu}}},\ }\href@noop {} {\bibfield  {journal} {\bibinfo  {journal} {Nat.
  Commun.}\ }\textbf {\bibinfo {volume} {8}},\ \bibinfo {pages} {14956}
  (\bibinfo {year} {2017})}\BibitemShut {NoStop}%
\bibitem [{\citenamefont {Xiao}\ \emph {et~al.}(2018)\citenamefont {Xiao},
  \citenamefont {Zhu}, \citenamefont {Wang}, \citenamefont {Feng},
  \citenamefont {Hu}, \citenamefont {Dasgupta}, \citenamefont {Han},
  \citenamefont {Wang}, \citenamefont {Muller}, \citenamefont {Martin},
  \citenamefont {Hu},\ and\ \citenamefont {Zhang}}]{xiao2018intrinsic}%
  \BibitemOpen
  \bibfield  {author} {\bibinfo {author} {\bibfnamefont {J.}~\bibnamefont
  {Xiao}}, \bibinfo {author} {\bibfnamefont {H.}~\bibnamefont {Zhu}}, \bibinfo
  {author} {\bibfnamefont {Y.}~\bibnamefont {Wang}}, \bibinfo {author}
  {\bibfnamefont {W.}~\bibnamefont {Feng}}, \bibinfo {author} {\bibfnamefont
  {Y.}~\bibnamefont {Hu}}, \bibinfo {author} {\bibfnamefont {A.}~\bibnamefont
  {Dasgupta}}, \bibinfo {author} {\bibfnamefont {Y.}~\bibnamefont {Han}},
  \bibinfo {author} {\bibfnamefont {Y.}~\bibnamefont {Wang}}, \bibinfo {author}
  {\bibfnamefont {D.~A.}\ \bibnamefont {Muller}}, \bibinfo {author}
  {\bibfnamefont {L.~W.}\ \bibnamefont {Martin}}, \bibinfo {author}
  {\bibfnamefont {P.~A.}\ \bibnamefont {Hu}},\ and\ \bibinfo {author}
  {\bibfnamefont {X.}~\bibnamefont {Zhang}},\ }\href
  {https://doi.org/10.1103/PhysRevLett.120.227601} {\bibfield  {journal}
  {\bibinfo  {journal} {Phys. Rev. Lett.}\ }\textbf {\bibinfo {volume} {120}},\
  \bibinfo {pages} {227601} (\bibinfo {year} {2018})}\BibitemShut {NoStop}%
\bibitem [{\citenamefont {{Zhao}}\ \emph {et~al.}(2018)\citenamefont {{Zhao}},
  \citenamefont {{Ma}}, \citenamefont {{Lv}}, \citenamefont {{Li}},
  \citenamefont {{Huang}},\ and\ \citenamefont {{Dai}}}]{zhao2018two}%
  \BibitemOpen
  \bibfield  {author} {\bibinfo {author} {\bibfnamefont {P.}~\bibnamefont
  {{Zhao}}}, \bibinfo {author} {\bibfnamefont {Y.}~\bibnamefont {{Ma}}},
  \bibinfo {author} {\bibfnamefont {X.}~\bibnamefont {{Lv}}}, \bibinfo {author}
  {\bibfnamefont {M.}~\bibnamefont {{Li}}}, \bibinfo {author} {\bibfnamefont
  {B.}~\bibnamefont {{Huang}}},\ and\ \bibinfo {author} {\bibfnamefont
  {Y.}~\bibnamefont {{Dai}}},\ }\href@noop {} {\bibfield  {journal} {\bibinfo
  {journal} {Nano Energy}\ }\textbf {\bibinfo {volume} {51}},\ \bibinfo {pages}
  {533} (\bibinfo {year} {2018})}\BibitemShut {NoStop}%
\bibitem [{\citenamefont {{Fu}}\ \emph {et~al.}(2018)\citenamefont {{Fu}},
  \citenamefont {{Sun}}, \citenamefont {{Luo}}, \citenamefont {{Li}},
  \citenamefont {{Hu}},\ and\ \citenamefont {{Yang}}}]{fu2018intrinsic}%
  \BibitemOpen
  \bibfield  {author} {\bibinfo {author} {\bibfnamefont {C.-F.}\ \bibnamefont
  {{Fu}}}, \bibinfo {author} {\bibfnamefont {J.}~\bibnamefont {{Sun}}},
  \bibinfo {author} {\bibfnamefont {Q.}~\bibnamefont {{Luo}}}, \bibinfo
  {author} {\bibfnamefont {X.}~\bibnamefont {{Li}}}, \bibinfo {author}
  {\bibfnamefont {W.}~\bibnamefont {{Hu}}},\ and\ \bibinfo {author}
  {\bibfnamefont {J.}~\bibnamefont {{Yang}}},\ }\href@noop {} {\bibfield
  {journal} {\bibinfo  {journal} {Nano Lett.}\ }\textbf {\bibinfo {volume}
  {18}},\ \bibinfo {pages} {6312} (\bibinfo {year} {2018})}\BibitemShut
  {NoStop}%
\bibitem [{\citenamefont {{Zhao}}\ \emph {et~al.}(2016)\citenamefont {{Zhao}},
  \citenamefont {{Baibuz}}, \citenamefont {{Vernieres}}, \citenamefont
  {{Grammatikopoulos}}, \citenamefont {{Jansson}}, \citenamefont {{Nagel}},
  \citenamefont {{Steinhauer}}, \citenamefont {{Sowwan}}, \citenamefont
  {{Kuronen}}, \citenamefont {{Nordlund}},\ and\ \citenamefont
  {{Djurabekova}}}]{zhao2016formation}%
  \BibitemOpen
  \bibfield  {author} {\bibinfo {author} {\bibfnamefont {J.}~\bibnamefont
  {{Zhao}}}, \bibinfo {author} {\bibfnamefont {E.}~\bibnamefont {{Baibuz}}},
  \bibinfo {author} {\bibfnamefont {J.}~\bibnamefont {{Vernieres}}}, \bibinfo
  {author} {\bibfnamefont {P.}~\bibnamefont {{Grammatikopoulos}}}, \bibinfo
  {author} {\bibfnamefont {V.}~\bibnamefont {{Jansson}}}, \bibinfo {author}
  {\bibfnamefont {M.}~\bibnamefont {{Nagel}}}, \bibinfo {author} {\bibfnamefont
  {S.}~\bibnamefont {{Steinhauer}}}, \bibinfo {author} {\bibfnamefont
  {M.}~\bibnamefont {{Sowwan}}}, \bibinfo {author} {\bibfnamefont
  {A.}~\bibnamefont {{Kuronen}}}, \bibinfo {author} {\bibfnamefont
  {K.}~\bibnamefont {{Nordlund}}},\ and\ \bibinfo {author} {\bibfnamefont
  {F.}~\bibnamefont {{Djurabekova}}},\ }\href@noop {} {\bibfield  {journal}
  {\bibinfo  {journal} {ACS Nano}\ }\textbf {\bibinfo {volume} {10}},\ \bibinfo
  {pages} {4684} (\bibinfo {year} {2016})}\BibitemShut {NoStop}%
\bibitem [{\citenamefont {{Vernieres}}\ \emph {et~al.}(2019)\citenamefont
  {{Vernieres}}, \citenamefont {{Steinhauer}}, \citenamefont {{Zhao}},
  \citenamefont {{Grammatikopoulos}}, \citenamefont {{Ferrando}}, \citenamefont
  {{Nordlund}}, \citenamefont {{Djurabekova}},\ and\ \citenamefont
  {{Sowwan}}}]{vernieres2019site}%
  \BibitemOpen
  \bibfield  {author} {\bibinfo {author} {\bibfnamefont {J.}~\bibnamefont
  {{Vernieres}}}, \bibinfo {author} {\bibfnamefont {S.}~\bibnamefont
  {{Steinhauer}}}, \bibinfo {author} {\bibfnamefont {J.}~\bibnamefont
  {{Zhao}}}, \bibinfo {author} {\bibfnamefont {P.}~\bibnamefont
  {{Grammatikopoulos}}}, \bibinfo {author} {\bibfnamefont {R.}~\bibnamefont
  {{Ferrando}}}, \bibinfo {author} {\bibfnamefont {K.}~\bibnamefont
  {{Nordlund}}}, \bibinfo {author} {\bibfnamefont {F.}~\bibnamefont
  {{Djurabekova}}},\ and\ \bibinfo {author} {\bibfnamefont {M.}~\bibnamefont
  {{Sowwan}}},\ }\href@noop {} {\bibfield  {journal} {\bibinfo  {journal} {Adv.
  Sci.}\ }\textbf {\bibinfo {volume} {6}},\ \bibinfo {pages} {1900447}
  (\bibinfo {year} {2019})}\BibitemShut {NoStop}%
\bibitem [{\citenamefont {Kresse}\ and\ \citenamefont
  {Hafner}(1993)}]{kresse1993ab}%
  \BibitemOpen
  \bibfield  {author} {\bibinfo {author} {\bibfnamefont {G.}~\bibnamefont
  {Kresse}}\ and\ \bibinfo {author} {\bibfnamefont {J.}~\bibnamefont
  {Hafner}},\ }\href@noop {} {\bibfield  {journal} {\bibinfo  {journal} {Phys.
  Rev. B}\ }\textbf {\bibinfo {volume} {47}},\ \bibinfo {pages} {558} (\bibinfo
  {year} {1993})}\BibitemShut {NoStop}%
\bibitem [{\citenamefont {Kresse}\ and\ \citenamefont
  {Furthm{\"u}ller}(1996)}]{kresse1996efficiency}%
  \BibitemOpen
  \bibfield  {author} {\bibinfo {author} {\bibfnamefont {G.}~\bibnamefont
  {Kresse}}\ and\ \bibinfo {author} {\bibfnamefont {J.}~\bibnamefont
  {Furthm{\"u}ller}},\ }\href@noop {} {\bibfield  {journal} {\bibinfo
  {journal} {Comput. Mater. Sci.}\ }\textbf {\bibinfo {volume} {6}},\ \bibinfo
  {pages} {15} (\bibinfo {year} {1996})}\BibitemShut {NoStop}%
\bibitem [{\citenamefont {Blochl}(1994)}]{PAW}%
  \BibitemOpen
  \bibfield  {author} {\bibinfo {author} {\bibfnamefont {P.~E.}\ \bibnamefont
  {Blochl}},\ }\href@noop {} {\bibfield  {journal} {\bibinfo  {journal} {Phys.
  Rev. B}\ }\textbf {\bibinfo {volume} {50}},\ \bibinfo {pages} {17953}
  (\bibinfo {year} {1994})}\BibitemShut {NoStop}%
\bibitem [{\citenamefont {Grimme}\ \emph {et~al.}(2010)\citenamefont {Grimme},
  \citenamefont {Antony}, \citenamefont {Ehrlich},\ and\ \citenamefont
  {Krieg}}]{DFTVDW}%
  \BibitemOpen
  \bibfield  {author} {\bibinfo {author} {\bibfnamefont {S.}~\bibnamefont
  {Grimme}}, \bibinfo {author} {\bibfnamefont {J.}~\bibnamefont {Antony}},
  \bibinfo {author} {\bibfnamefont {S.}~\bibnamefont {Ehrlich}},\ and\ \bibinfo
  {author} {\bibfnamefont {H.}~\bibnamefont {Krieg}},\ }\href@noop {}
  {\bibfield  {journal} {\bibinfo  {journal} {J. Chem. Phys.}\ }\textbf
  {\bibinfo {volume} {132}},\ \bibinfo {pages} {154104} (\bibinfo {year}
  {2010})}\BibitemShut {NoStop}%
\bibitem [{\citenamefont {Perdew}\ \emph {et~al.}(1996)\citenamefont {Perdew},
  \citenamefont {Burke},\ and\ \citenamefont
  {Ernzerhof}}]{perdew1996generalized}%
  \BibitemOpen
  \bibfield  {author} {\bibinfo {author} {\bibfnamefont {J.~P.}\ \bibnamefont
  {Perdew}}, \bibinfo {author} {\bibfnamefont {K.}~\bibnamefont {Burke}},\ and\
  \bibinfo {author} {\bibfnamefont {M.}~\bibnamefont {Ernzerhof}},\ }\href@noop
  {} {\bibfield  {journal} {\bibinfo  {journal} {Phys. Rev. Lett.}\ }\textbf
  {\bibinfo {volume} {77}},\ \bibinfo {pages} {3865} (\bibinfo {year}
  {1996})}\BibitemShut {NoStop}%
\bibitem [{\citenamefont {{Cohen}}\ \emph {et~al.}(2008)\citenamefont
  {{Cohen}}, \citenamefont {{Mori-Sánchez}},\ and\ \citenamefont
  {{Yang}}}]{cohen2008insights}%
  \BibitemOpen
  \bibfield  {author} {\bibinfo {author} {\bibfnamefont {A.~J.}\ \bibnamefont
  {{Cohen}}}, \bibinfo {author} {\bibfnamefont {P.}~\bibnamefont
  {{Mori-Sánchez}}},\ and\ \bibinfo {author} {\bibfnamefont {W.}~\bibnamefont
  {{Yang}}},\ }\href@noop {} {\bibfield  {journal} {\bibinfo  {journal}
  {Science}\ }\textbf {\bibinfo {volume} {321}},\ \bibinfo {pages} {792}
  (\bibinfo {year} {2008})}\BibitemShut {NoStop}%
\bibitem [{\citenamefont {{Momeni}}\ \emph {et~al.}(2020)\citenamefont
  {{Momeni}}, \citenamefont {{Ji}}, \citenamefont {{Wang}}, \citenamefont
  {{Paul}}, \citenamefont {{Neshani}}, \citenamefont {{Yilmaz}}, \citenamefont
  {{Shin}}, \citenamefont {{Zhang}}, \citenamefont {{Jiang}}, \citenamefont
  {{Park}}, \citenamefont {{Sinnott}}, \citenamefont {van {Duin}},
  \citenamefont {{Crespi}},\ and\ \citenamefont {qing
  {Chen}}}]{momeni2020multiscale}%
  \BibitemOpen
  \bibfield  {author} {\bibinfo {author} {\bibfnamefont {K.}~\bibnamefont
  {{Momeni}}}, \bibinfo {author} {\bibfnamefont {Y.}~\bibnamefont {{Ji}}},
  \bibinfo {author} {\bibfnamefont {Y.}~\bibnamefont {{Wang}}}, \bibinfo
  {author} {\bibfnamefont {S.}~\bibnamefont {{Paul}}}, \bibinfo {author}
  {\bibfnamefont {S.}~\bibnamefont {{Neshani}}}, \bibinfo {author}
  {\bibfnamefont {D.~E.}\ \bibnamefont {{Yilmaz}}}, \bibinfo {author}
  {\bibfnamefont {Y.~K.}\ \bibnamefont {{Shin}}}, \bibinfo {author}
  {\bibfnamefont {D.}~\bibnamefont {{Zhang}}}, \bibinfo {author} {\bibfnamefont
  {J.~W.}\ \bibnamefont {{Jiang}}}, \bibinfo {author} {\bibfnamefont {H.~S.}\
  \bibnamefont {{Park}}}, \bibinfo {author} {\bibfnamefont {S.~B.}\
  \bibnamefont {{Sinnott}}}, \bibinfo {author} {\bibfnamefont {A.}~\bibnamefont
  {van {Duin}}}, \bibinfo {author} {\bibfnamefont {V.~H.}\ \bibnamefont
  {{Crespi}}},\ and\ \bibinfo {author} {\bibfnamefont {L.}~\bibnamefont {qing
  {Chen}}},\ }\href@noop {} {\bibfield  {journal} {\bibinfo  {journal} {npj
  Comput. Mater.}\ }\textbf {\bibinfo {volume} {6}},\ \bibinfo {pages} {1}
  (\bibinfo {year} {2020})}\BibitemShut {NoStop}%
\bibitem [{\citenamefont {{Jonsson}}\ \emph {et~al.}(1998)\citenamefont
  {{Jonsson}}, \citenamefont {{Mills}},\ and\ \citenamefont
  {{Jacobsen}}}]{jonsson1998nudged}%
  \BibitemOpen
  \bibfield  {author} {\bibinfo {author} {\bibfnamefont {H.}~\bibnamefont
  {{Jonsson}}}, \bibinfo {author} {\bibfnamefont {G.}~\bibnamefont {{Mills}}},\
  and\ \bibinfo {author} {\bibfnamefont {K.~W.}\ \bibnamefont {{Jacobsen}}},\
  }in\ \href@noop {} {\emph {\bibinfo {booktitle} {Classical and Quantum
  Dynamics in Condensed Phase Simulations}}},\ Vol.\ \bibinfo {volume} {385}\
  (\bibinfo {year} {1998})\ pp.\ \bibinfo {pages} {385--404}\BibitemShut
  {NoStop}%
\bibitem [{\citenamefont {{Henkelman}}\ \emph {et~al.}(2000)\citenamefont
  {{Henkelman}}, \citenamefont {{Uberuaga}},\ and\ \citenamefont
  {{Jónsson}}}]{henkelman2000a}%
  \BibitemOpen
  \bibfield  {author} {\bibinfo {author} {\bibfnamefont {G.}~\bibnamefont
  {{Henkelman}}}, \bibinfo {author} {\bibfnamefont {B.~P.}\ \bibnamefont
  {{Uberuaga}}},\ and\ \bibinfo {author} {\bibfnamefont {H.}~\bibnamefont
  {{Jónsson}}},\ }\href@noop {} {\bibfield  {journal} {\bibinfo  {journal} {J.
  Chem. Phys.}\ }\textbf {\bibinfo {volume} {113}},\ \bibinfo {pages} {9901}
  (\bibinfo {year} {2000})}\BibitemShut {NoStop}%
\bibitem [{\citenamefont {{Henkelman}}\ and\ \citenamefont
  {{Jónsson}}(2000)}]{henkelman2000improved}%
  \BibitemOpen
  \bibfield  {author} {\bibinfo {author} {\bibfnamefont {G.~A.}\ \bibnamefont
  {{Henkelman}}}\ and\ \bibinfo {author} {\bibfnamefont {H.}~\bibnamefont
  {{Jónsson}}},\ }\href@noop {} {\bibfield  {journal} {\bibinfo  {journal} {J.
  Chem. Phys.}\ }\textbf {\bibinfo {volume} {113}},\ \bibinfo {pages} {9978}
  (\bibinfo {year} {2000})}\BibitemShut {NoStop}%
\bibitem [{\citenamefont {{Sheppard}}\ \emph {et~al.}(2012)\citenamefont
  {{Sheppard}}, \citenamefont {{Xiao}}, \citenamefont {{Chemelewski}},
  \citenamefont {{Johnson}},\ and\ \citenamefont {{Henkelman}}}]{SSNEB}%
  \BibitemOpen
  \bibfield  {author} {\bibinfo {author} {\bibfnamefont {D.}~\bibnamefont
  {{Sheppard}}}, \bibinfo {author} {\bibfnamefont {P.}~\bibnamefont {{Xiao}}},
  \bibinfo {author} {\bibfnamefont {W.}~\bibnamefont {{Chemelewski}}}, \bibinfo
  {author} {\bibfnamefont {D.~D.}\ \bibnamefont {{Johnson}}},\ and\ \bibinfo
  {author} {\bibfnamefont {G.}~\bibnamefont {{Henkelman}}},\ }\href@noop {}
  {\bibfield  {journal} {\bibinfo  {journal} {J. Chem. Phys.}\ }\textbf
  {\bibinfo {volume} {136}},\ \bibinfo {pages} {74103} (\bibinfo {year}
  {2012})}\BibitemShut {NoStop}%
\bibitem [{\citenamefont {{Larsen}}\ \emph {et~al.}(2017)\citenamefont
  {{Larsen}}, \citenamefont {{Mortensen}}, \citenamefont {{Blomqvist}},
  \citenamefont {{Castelli}}, \citenamefont {{Christensen}}, \citenamefont
  {{Dulak}}, \citenamefont {{Friis}}, \citenamefont {{Groves}}, \citenamefont
  {{Hammer}}, \citenamefont {{Hargus}}, \citenamefont {{Hermes}}, \citenamefont
  {{Jennings}}, \citenamefont {{Jensen}}, \citenamefont {{Kermode}},
  \citenamefont {{Kitchin}}, \citenamefont {{Kolsbjerg}}, \citenamefont
  {{Kubal}}, \citenamefont {{Kaasbjerg}}, \citenamefont {{Lysgaard}},
  \citenamefont {{Maronsson}}, \citenamefont {{Maxson}}, \citenamefont
  {{Olsen}}, \citenamefont {{Pastewka}}, \citenamefont {{Peterson}},
  \citenamefont {{Rostgaard}}, \citenamefont {{Schiøtz}}, \citenamefont
  {{Schütt}}, \citenamefont {{Strange}}, \citenamefont {{Thygesen}},
  \citenamefont {{Vegge}}, \citenamefont {{Vilhelmsen}}, \citenamefont
  {{Walter}}, \citenamefont {{Zeng}},\ and\ \citenamefont {{Jacobsen}}}]{ASE}%
  \BibitemOpen
  \bibfield  {author} {\bibinfo {author} {\bibfnamefont {A.~H.}\ \bibnamefont
  {{Larsen}}}, \bibinfo {author} {\bibfnamefont {J.~J.}\ \bibnamefont
  {{Mortensen}}}, \bibinfo {author} {\bibfnamefont {J.}~\bibnamefont
  {{Blomqvist}}}, \bibinfo {author} {\bibfnamefont {I.~E.}\ \bibnamefont
  {{Castelli}}}, \bibinfo {author} {\bibfnamefont {R.}~\bibnamefont
  {{Christensen}}}, \bibinfo {author} {\bibfnamefont {M.}~\bibnamefont
  {{Dulak}}}, \bibinfo {author} {\bibfnamefont {J.}~\bibnamefont {{Friis}}},
  \bibinfo {author} {\bibfnamefont {M.~N.}\ \bibnamefont {{Groves}}}, \bibinfo
  {author} {\bibfnamefont {B.}~\bibnamefont {{Hammer}}}, \bibinfo {author}
  {\bibfnamefont {C.}~\bibnamefont {{Hargus}}}, \bibinfo {author}
  {\bibfnamefont {E.~D.}\ \bibnamefont {{Hermes}}}, \bibinfo {author}
  {\bibfnamefont {P.~C.}\ \bibnamefont {{Jennings}}}, \bibinfo {author}
  {\bibfnamefont {P.~B.}\ \bibnamefont {{Jensen}}}, \bibinfo {author}
  {\bibfnamefont {J.}~\bibnamefont {{Kermode}}}, \bibinfo {author}
  {\bibfnamefont {J.~R.}\ \bibnamefont {{Kitchin}}}, \bibinfo {author}
  {\bibfnamefont {E.~L.}\ \bibnamefont {{Kolsbjerg}}}, \bibinfo {author}
  {\bibfnamefont {J.}~\bibnamefont {{Kubal}}}, \bibinfo {author} {\bibfnamefont
  {K.}~\bibnamefont {{Kaasbjerg}}}, \bibinfo {author} {\bibfnamefont
  {S.}~\bibnamefont {{Lysgaard}}}, \bibinfo {author} {\bibfnamefont {J.~B.}\
  \bibnamefont {{Maronsson}}}, \bibinfo {author} {\bibfnamefont
  {T.}~\bibnamefont {{Maxson}}}, \bibinfo {author} {\bibfnamefont
  {T.}~\bibnamefont {{Olsen}}}, \bibinfo {author} {\bibfnamefont
  {L.}~\bibnamefont {{Pastewka}}}, \bibinfo {author} {\bibfnamefont
  {A.}~\bibnamefont {{Peterson}}}, \bibinfo {author} {\bibfnamefont
  {C.}~\bibnamefont {{Rostgaard}}}, \bibinfo {author} {\bibfnamefont
  {J.}~\bibnamefont {{Schiøtz}}}, \bibinfo {author} {\bibfnamefont
  {O.}~\bibnamefont {{Schütt}}}, \bibinfo {author} {\bibfnamefont
  {M.}~\bibnamefont {{Strange}}}, \bibinfo {author} {\bibfnamefont {K.~S.}\
  \bibnamefont {{Thygesen}}}, \bibinfo {author} {\bibfnamefont
  {T.}~\bibnamefont {{Vegge}}}, \bibinfo {author} {\bibfnamefont
  {L.}~\bibnamefont {{Vilhelmsen}}}, \bibinfo {author} {\bibfnamefont {M.~N.}\
  \bibnamefont {{Walter}}}, \bibinfo {author} {\bibfnamefont {Z.}~\bibnamefont
  {{Zeng}}},\ and\ \bibinfo {author} {\bibfnamefont {K.~W.}\ \bibnamefont
  {{Jacobsen}}},\ }\href@noop {} {\bibfield  {journal} {\bibinfo  {journal} {J.
  Phys.: Condens. Matter}\ }\textbf {\bibinfo {volume} {29}},\ \bibinfo {pages}
  {273002} (\bibinfo {year} {2017})}\BibitemShut {NoStop}%
\bibitem [{TSA()}]{TSASE}%
  \BibitemOpen
  \href {http://theory.cm.utexas.edu/tsase/} {\bibinfo {title} {About tsase:
  Transition state library for ase,
  http://theory.cm.utexas.edu/tsase/}}\BibitemShut {NoStop}%
\bibitem [{\citenamefont {{Stukowski}}(2010)}]{stukowski2010visualization}%
  \BibitemOpen
  \bibfield  {author} {\bibinfo {author} {\bibfnamefont {A.}~\bibnamefont
  {{Stukowski}}},\ }\href@noop {} {\bibfield  {journal} {\bibinfo  {journal}
  {Modell. Simul. Mater. Sci. Eng.}\ }\textbf {\bibinfo {volume} {18}},\
  \bibinfo {pages} {15012} (\bibinfo {year} {2010})}\BibitemShut {NoStop}%
\bibitem [{\citenamefont {Bart\'ok}\ \emph {et~al.}(2010)\citenamefont
  {Bart\'ok}, \citenamefont {Payne}, \citenamefont {Kondor},\ and\
  \citenamefont {Cs\'anyi}}]{GAP}%
  \BibitemOpen
  \bibfield  {author} {\bibinfo {author} {\bibfnamefont {A.~P.}\ \bibnamefont
  {Bart\'ok}}, \bibinfo {author} {\bibfnamefont {M.~C.}\ \bibnamefont {Payne}},
  \bibinfo {author} {\bibfnamefont {R.}~\bibnamefont {Kondor}},\ and\ \bibinfo
  {author} {\bibfnamefont {G.}~\bibnamefont {Cs\'anyi}},\ }\href
  {https://doi.org/10.1103/PhysRevLett.104.136403} {\bibfield  {journal}
  {\bibinfo  {journal} {Phys. Rev. Lett.}\ }\textbf {\bibinfo {volume} {104}},\
  \bibinfo {pages} {136403} (\bibinfo {year} {2010})}\BibitemShut {NoStop}%
\bibitem [{\citenamefont {Bart\'ok}\ \emph {et~al.}(2013)\citenamefont
  {Bart\'ok}, \citenamefont {Kondor},\ and\ \citenamefont {Cs\'anyi}}]{SOAP}%
  \BibitemOpen
  \bibfield  {author} {\bibinfo {author} {\bibfnamefont {A.~P.}\ \bibnamefont
  {Bart\'ok}}, \bibinfo {author} {\bibfnamefont {R.}~\bibnamefont {Kondor}},\
  and\ \bibinfo {author} {\bibfnamefont {G.}~\bibnamefont {Cs\'anyi}},\ }\href
  {https://doi.org/10.1103/PhysRevB.87.184115} {\bibfield  {journal} {\bibinfo
  {journal} {Phys. Rev. B}\ }\textbf {\bibinfo {volume} {87}},\ \bibinfo
  {pages} {184115} (\bibinfo {year} {2013})}\BibitemShut {NoStop}%
\bibitem [{\citenamefont {{Liu}}\ \emph {et~al.}(2020)\citenamefont {{Liu}},
  \citenamefont {{Yang}}, \citenamefont {{Xin}}, \citenamefont {{Liu}},
  \citenamefont {{Csányi}},\ and\ \citenamefont {{Cao}}}]{liu2020machine}%
  \BibitemOpen
  \bibfield  {author} {\bibinfo {author} {\bibfnamefont {Y.-B.}\ \bibnamefont
  {{Liu}}}, \bibinfo {author} {\bibfnamefont {J.-Y.}\ \bibnamefont {{Yang}}},
  \bibinfo {author} {\bibfnamefont {G.-M.}\ \bibnamefont {{Xin}}}, \bibinfo
  {author} {\bibfnamefont {L.-H.}\ \bibnamefont {{Liu}}}, \bibinfo {author}
  {\bibfnamefont {G.}~\bibnamefont {{Csányi}}},\ and\ \bibinfo {author}
  {\bibfnamefont {B.-Y.}\ \bibnamefont {{Cao}}},\ }\href@noop {} {\bibfield
  {journal} {\bibinfo  {journal} {J. Chem. Phys.}\ }\textbf {\bibinfo {volume}
  {153}},\ \bibinfo {pages} {144501} (\bibinfo {year} {2020})}\BibitemShut
  {NoStop}%
\bibitem [{\citenamefont {{Li}}\ \emph {et~al.}(2020)\citenamefont {{Li}},
  \citenamefont {{Liu}}, \citenamefont {{Rohskopf}}, \citenamefont {{Gordiz}},
  \citenamefont {{Henry}}, \citenamefont {{Lee}},\ and\ \citenamefont
  {{Luo}}}]{li2020a}%
  \BibitemOpen
  \bibfield  {author} {\bibinfo {author} {\bibfnamefont {R.}~\bibnamefont
  {{Li}}}, \bibinfo {author} {\bibfnamefont {Z.}~\bibnamefont {{Liu}}},
  \bibinfo {author} {\bibfnamefont {A.}~\bibnamefont {{Rohskopf}}}, \bibinfo
  {author} {\bibfnamefont {K.}~\bibnamefont {{Gordiz}}}, \bibinfo {author}
  {\bibfnamefont {A.}~\bibnamefont {{Henry}}}, \bibinfo {author} {\bibfnamefont
  {E.}~\bibnamefont {{Lee}}},\ and\ \bibinfo {author} {\bibfnamefont
  {T.}~\bibnamefont {{Luo}}},\ }\href@noop {} {\bibfield  {journal} {\bibinfo
  {journal} {Appl. Phys. Lett.}\ }\textbf {\bibinfo {volume} {117}},\ \bibinfo
  {pages} {152102} (\bibinfo {year} {2020})}\BibitemShut {NoStop}%
\bibitem [{\citenamefont {Pozdnyakov}\ \emph {et~al.}(2020)\citenamefont
  {Pozdnyakov}, \citenamefont {Willatt}, \citenamefont {Bart\'ok},
  \citenamefont {Ortner}, \citenamefont {Cs\'anyi},\ and\ \citenamefont
  {Ceriotti}}]{NEW2B}%
  \BibitemOpen
  \bibfield  {author} {\bibinfo {author} {\bibfnamefont {S.~N.}\ \bibnamefont
  {Pozdnyakov}}, \bibinfo {author} {\bibfnamefont {M.~J.}\ \bibnamefont
  {Willatt}}, \bibinfo {author} {\bibfnamefont {A.~P.}\ \bibnamefont
  {Bart\'ok}}, \bibinfo {author} {\bibfnamefont {C.}~\bibnamefont {Ortner}},
  \bibinfo {author} {\bibfnamefont {G.}~\bibnamefont {Cs\'anyi}},\ and\
  \bibinfo {author} {\bibfnamefont {M.}~\bibnamefont {Ceriotti}},\ }\href
  {https://doi.org/10.1103/PhysRevLett.125.166001} {\bibfield  {journal}
  {\bibinfo  {journal} {Phys. Rev. Lett.}\ }\textbf {\bibinfo {volume} {125}},\
  \bibinfo {pages} {166001} (\bibinfo {year} {2020})}\BibitemShut {NoStop}%
\bibitem [{\citenamefont {{Bartók}}\ and\ \citenamefont
  {{Csányi}}(2015)}]{GAP2015}%
  \BibitemOpen
  \bibfield  {author} {\bibinfo {author} {\bibfnamefont {A.~P.}\ \bibnamefont
  {{Bartók}}}\ and\ \bibinfo {author} {\bibfnamefont {G.}~\bibnamefont
  {{Csányi}}},\ }\href@noop {} {\bibfield  {journal} {\bibinfo  {journal}
  {Int. J. Quantum Chem.}\ }\textbf {\bibinfo {volume} {115}},\ \bibinfo
  {pages} {1051} (\bibinfo {year} {2015})}\BibitemShut {NoStop}%
\bibitem [{QUI()}]{QUIP}%
  \BibitemOpen
  \href {https://github.com/libAtoms/QUIP} {\bibinfo {title} {Quip - quantum
  mechanics and interatomic potentials,
  https://github.com/libatoms/quip}}\BibitemShut {NoStop}%
\bibitem [{\citenamefont {{Scott}}(2007)}]{Scott2007ferro}%
  \BibitemOpen
  \bibfield  {author} {\bibinfo {author} {\bibfnamefont {J.~F.}\ \bibnamefont
  {{Scott}}},\ }\href@noop {} {\bibfield  {journal} {\bibinfo  {journal}
  {Science}\ }\textbf {\bibinfo {volume} {315}},\ \bibinfo {pages} {954}
  (\bibinfo {year} {2007})}\BibitemShut {NoStop}%
\bibitem [{\citenamefont {{Dawber}}\ \emph {et~al.}(2005)\citenamefont
  {{Dawber}}, \citenamefont {{Rabe}},\ and\ \citenamefont
  {{Scott}}}]{dawber2005physics}%
  \BibitemOpen
  \bibfield  {author} {\bibinfo {author} {\bibfnamefont {M.}~\bibnamefont
  {{Dawber}}}, \bibinfo {author} {\bibfnamefont {K.~M.}\ \bibnamefont
  {{Rabe}}},\ and\ \bibinfo {author} {\bibfnamefont {J.~F.}\ \bibnamefont
  {{Scott}}},\ }\href@noop {} {\bibfield  {journal} {\bibinfo  {journal} {Rev.
  Mod. Phys.}\ }\textbf {\bibinfo {volume} {77}},\ \bibinfo {pages} {1083}
  (\bibinfo {year} {2005})}\BibitemShut {NoStop}%
\bibitem [{\citenamefont {Laidler}(1984)}]{Arrheniuseq}%
  \BibitemOpen
  \bibfield  {author} {\bibinfo {author} {\bibfnamefont {K.~J.}\ \bibnamefont
  {Laidler}},\ }\href@noop {} {\bibfield  {journal} {\bibinfo  {journal} {J.
  Chem. Educ.}\ }\textbf {\bibinfo {volume} {61}},\ \bibinfo {pages} {494}
  (\bibinfo {year} {1984})}\BibitemShut {NoStop}%
\bibitem [{GAP()}]{GAP2021}%
  \BibitemOpen
  \href@noop {} {}\bibinfo {howpublished} {private communication}\BibitemShut
  {NoStop}%
\bibitem [{\citenamefont {{He}}\ \emph {et~al.}(2006)\citenamefont {{He}},
  \citenamefont {{Blanco}},\ and\ \citenamefont {{Pandey}}}]{he2006electronic}%
  \BibitemOpen
  \bibfield  {author} {\bibinfo {author} {\bibfnamefont {H.}~\bibnamefont
  {{He}}}, \bibinfo {author} {\bibfnamefont {M.~A.}\ \bibnamefont {{Blanco}}},\
  and\ \bibinfo {author} {\bibfnamefont {R.}~\bibnamefont {{Pandey}}},\
  }\href@noop {} {\bibfield  {journal} {\bibinfo  {journal} {Appl. Phys.
  Lett.}\ }\textbf {\bibinfo {volume} {88}},\ \bibinfo {pages} {261904}
  (\bibinfo {year} {2006})}\BibitemShut {NoStop}%
\bibitem [{\citenamefont {{Henkelman}}\ \emph {et~al.}(2006)\citenamefont
  {{Henkelman}}, \citenamefont {{Arnaldsson}},\ and\ \citenamefont
  {{Jónsson}}}]{henkelman2006a}%
  \BibitemOpen
  \bibfield  {author} {\bibinfo {author} {\bibfnamefont {G.~A.}\ \bibnamefont
  {{Henkelman}}}, \bibinfo {author} {\bibfnamefont {A.}~\bibnamefont
  {{Arnaldsson}}},\ and\ \bibinfo {author} {\bibfnamefont {H.}~\bibnamefont
  {{Jónsson}}},\ }\href@noop {} {\bibfield  {journal} {\bibinfo  {journal}
  {Comput. Mater. Sci.}\ }\textbf {\bibinfo {volume} {36}},\ \bibinfo {pages}
  {354} (\bibinfo {year} {2006})}\BibitemShut {NoStop}%
\bibitem [{\citenamefont {Yu}\ and\ \citenamefont
  {Trinkle}(2011)}]{yu2011accurate}%
  \BibitemOpen
  \bibfield  {author} {\bibinfo {author} {\bibfnamefont {M.}~\bibnamefont
  {Yu}}\ and\ \bibinfo {author} {\bibfnamefont {D.~R.}\ \bibnamefont
  {Trinkle}},\ }\href@noop {} {\bibfield  {journal} {\bibinfo  {journal} {J.
  Chem. Phys.}\ }\textbf {\bibinfo {volume} {134}},\ \bibinfo {pages} {064111}
  (\bibinfo {year} {2011})}\BibitemShut {NoStop}%
\bibitem [{\citenamefont {{Silvi}}\ and\ \citenamefont {{Savin}}(1994)}]{ELF}%
  \BibitemOpen
  \bibfield  {author} {\bibinfo {author} {\bibfnamefont {B.}~\bibnamefont
  {{Silvi}}}\ and\ \bibinfo {author} {\bibfnamefont {A.}~\bibnamefont
  {{Savin}}},\ }\href@noop {} {\bibfield  {journal} {\bibinfo  {journal}
  {Nature}\ }\textbf {\bibinfo {volume} {371}},\ \bibinfo {pages} {683}
  (\bibinfo {year} {1994})}\BibitemShut {NoStop}%
\end{thebibliography}%

\end{document}